\documentclass[12pt]{osu_phd}
\newcommand{\nn}{\nonumber}
\newcommand{\e}{\epsilon}
\newcommand{\beq}{\begin{equation} }
\newcommand{\eeq} {\end{equation} }
\newcommand{\bea}{\begin{eqnarray} }
\newcommand{\eea} {\end{eqnarray} }

\newcommand{\bed}{\begin{displaymath} }
\newcommand{\eed} {\end{displaymath} }

\usepackage{epsf}
\usepackage{cancel}
\usepackage{amssymb}
\usepackage{amsmath}
\usepackage{graphicx}
\usepackage{psfig}
\usepackage{makeidx}
\usepackage{ifthen}
\begin{document}

\ifx\href\undefined\else\hypersetup{linktocpage=true}\fi

%
%
%
\chapternumberon 
%
%
%
%

%
\title{HIDDEN SYMMETRIES AND \\ THEIR IMPLICATIONS FOR PARTICLE PHYSICS}
\author{KAI WANG\\~\\Bachelor of Science\\Zhejiang University
\\Hangzhou, China\\2000}
\date{July, 2004}
\degree{MASTER OF SCIENCE}

\maketitle

%
%
%
%
%
%

\thispagestyle{myheadings}              

\begin{center}
       HIDDEN SYMMETRIES AND \\ THEIR IMPLICATIONS FOR PARTICLE PHYSICS                                     
\end{center}

\hspace*{3.1em} Thesis Approved:

\vspace{20pt}

\begin{center}
Prof. K.S. Babu, {\footnotesize Thesis Advisor}\\

Prof. John Mintmire\\
Prof. Paul Westhaus\\
Prof. Al Carlozzi, {\footnotesize Dean of the Graduate College}


\end{center}
\pagebreak

%
%
%
%
\pagestyle{edheadings}
\vspace*{.605in}
\begin{center}
ACKNOWLEDGMENTS
\end{center}
\vspace*{0.19in}

There is usually one person who played the pivotal role as a
mentor in one's graduate career. To me, this person has been Prof.~K.S. Babu. As such an outstanding model building phenomenologist,
he has a distinguished view of connection between experiments and
theories. I wish to express my deepest gratitude to him for his constructive
guidance, constant encouragement, kindness and great patience. Without
his guidance and collaboration, I would not have been able to finish this
work. Many thanks are also due to Prof.~Mingxing Luo, who has led me into theoretical high energy physics many years ago;
Ilia Gogoladze, our former postdoctoral research associate, who has also been
a good teacher and a wonderful collaborator for years, and our current postdoctoral research associate,
Gerhart Seidl, whose encouragement and suggestion has been invaluable in completing this work.

I would also like to express my sincere appreciation to Prof.~Paul Westhaus for his assistance during my study
and my other committee member Prof.~John Mintmire. My appreciation extends to Prof.~Satya Nandi, Prof.~Jacques H.H. Perk,
Prof.~Aihua Xie, Prof.~Xincheng Xie, Gary Don Thacker, Susan
Cantrell,  Cindi Raymond, Karen S. Hopkins, and all our
faculty members and staff in the department. Moreover, I want to thank many of my physics
colleagues for their support and friendship, especially,  Ye
Xiong,  Zhenyue Zhu,  Brian Timmons,  Jianning Wang, Beining Nie,
Cyril Anoka, Abdelghafour Bachri,  Tsedenbaljir Enkhbat and Corneliu Marius Rujoiu.

I would like to thank the physics department at Oklahoma State University for years of support during my study and the U.S. Department of Energy
for the summer financial support in part by DOE
Grant \# DE-FG03-98ER-41076, DOE Grant \# DE-FG02-01ER-45684.

I wish to thank my parents for their constant support and encouragement.
Finally, I want to thank Qinan for her support and patiently
listening while I talked about my work.  She has made graduate
school which should have seemed interminable, instead sweet and
happy. I look forward to many more years of the same.


%
\textheight 8.333in
\tableofcontents
\listoftables
\listoffigures
%
\setlength{\textheight}{9.0in}
\resetcounters
\pagenumbering{arabic}
\chapter{\textsc{Introduction}}

\section{The Standard Model and Beyond}

The Standard Model (SM) of elementary particle physics is a chiral gauge theory that gives a successful description of
 strong, weak and electromagnetic interactions \cite{sm}. It has been highly successful in explaining all
experimental observations in the energy regime up to $M_{\rm E\rm W}\sim {\mathcal
O}(10^2 ~{\rm G\rm e\rm V})$. The theory is invariant under the gauge group
  $G_{\rm S\rm M}= SU(3)_C\times SU(2)_L\times U(1)_Y$. The $SU(3)_C$ Quantum Chromodynamics (QCD) describes
 the strong interaction which is supported by evidence from deep inelastic collision experiments. The $SU(2)_L\times U(1)_Y$ gauge
 symmetry corresponds to the Weinberg-Salam model of electroweak interaction which has been verified by a host of experiments,
 including the UA1/UA2 \cite{ua2} and LEP \cite{lep}.

In this thesis, we use the conventional notations for the SM matter fields. They are shown in Table \ref{sm}.
\begin{table}[ht]\label{sm}
 \begin{center}
  {\renewcommand{\arraystretch}{1.1}
 \begin{tabular}{c c c c c c c c }
   \hline
     \rule[5mm]{0mm}{0pt} & $Q$ & $u^c$ & $d^c$ & $\ell$ & $e^c$ & $H$ \\
  \hline
\rule[5mm]{0mm}{0pt}
   $SU(3)_{C}$ &  \bf{3} & $\bar{\bf{3}}$ & $\bar{\bf{3}}$ &
    \bf{1} &  \bf{1} &  \bf{1}  \\
    \rule[5mm]{0mm}{0pt}
  $ SU(2)_{L}$ &  \bf{2} &  \bf{1} &  \bf{1} &
    \bf{2} &  \bf{1} &  \bf{2}   \\
   \rule[5mm]{0mm}{0pt}
   $U(1)_Y$ & 1/6 & $-2/3$ & 1/3 & $-1/2$ & 1 & 1/2\\
    \hline
 \end{tabular}
 }
  \caption{Transformation properties of the SM fields under $G_\rm{SM}$}
 \end{center}
\end{table}

As a chiral theory, the left-handed and right-handed fermions have different transformation properties with respect to $G_\rm{SM}$.
Under $SU(2)_L$, the left-handed particles transform as the doublets \bea
Q=\left(\begin{array}{cc}u~~c~~t \\d~~s~~b \end{array}\right)_{L}~~{\rm a\rm n\rm d}~~
\ell=\left(\begin{array}{cc}\nu_e ~~\nu_\mu~~ \nu_\tau\\e ~~\mu~~ \tau\end{array}\right)_{L},\nonumber
\eea
while right-handed particles are $SU(2)_L$ singlets:
\beq (u^c,~c^c,~t^c), ~~(d^c,~s^c,~b^c),~~(e^c,~\mu^c,~\tau^c).\nn\eeq
The electroweak gauge symmetry is spontaneously broken via the Higgs mechanism by a scalar $SU(2)$ doublet \cite{higgs}
\bea
H=\left(\begin{array}{cc} H^+\\ H^0\end{array}\right).\nn\eea The masses of the quarks and leptons arise from Yukawa couplings from the lagrangian:
\beq\label{smyukawa} {\mathcal L}_{\rm S\rm M}=Y_u Q u^c H +Y_d Q d^c \bar{H} +Y_e \ell e^c \bar{H}+ {\rm h.\rm c.}, \eeq
where $\bar{H}$ is defined by $\bar{H}=i\sigma_2 H^\dag$ and $Y_u$, $Y_d$ and $Y_e$ are dimensionless coupling constants known as
Yukawa couplings. Note that the generation and color indices are contracted here.

Quantum correction to the Higgs boson mass induces the only quadratic divergence in the theory. For example, at the one-loop-level, the top quark
Yukawa couplings induces a quadratic divergence given by
\bed \Delta m^2_H= \frac{\lambda^2_t}{8\pi^2}\Lambda^2, \eed where the cutoff scale $\Lambda$
can be as large as $M_{\rm P\rm l}$ of order $\mathcal{O}(10^{19}~\rm{GeV})$  \cite{hierarchy}. If so,
the entire Higgs mechanism explanation of electroweak symmetry breaking would fail or require fine tuning of parameters.
In order to address this so-called gauge hierarchy problem \cite{hierarchy},
one would have to introduce new physics at the TeV scale. The most elegant solution
is known as supersymmetry (SUSY) \cite{susy}, where to each particle, there exists a SUSY partner with different spin. For instance, the superpartner
of the matter fermion top quark is a scalar known as \textit{stop} $\tilde{t}$. The quadratic divergence in the Higgs mass is now
 removed via the cancellation between top loop and stop loop. At the one-loop-level, it is
\beq \Delta m^2_H= \frac{\lambda^2_t}{8\pi^2}\Lambda^2+
\frac{\lambda^2_t}{8\pi^2}(m^2_{\tilde{t}}-\Lambda^2)\eeq
This cancellation is valid up to all loop corrections and is thus technically natural.

The minimal SUSY version of the SM is called the Minimal Supersymmetric Standard Model (MSSM) \cite{susy,susyreview} which is
described by the superpotential
\beq {\mathcal W}_{\rm M\rm S\rm S\rm M}= Y_u Q u^c H_u +Y_d Q d^c H_d +Y_e \ell e^c H_d+\mu H_u H_d, \eeq
where all the matter fields are now chiral superfields and two Higgs doublets\bea
H_{d}=\left(\begin{array}{c
c} H_{d}^0\\H_d^-\end{array}\right)~~{\rm a\rm n\rm d}~~ H_{u}=\left(\begin{array}{c
c} H_u^+\\H_u^0\end{array}\right) \nonumber
\eea are introduced since the superpotential must be holomorphic, i.e., the MSSM is a two-Higgs model \cite{susyreview}. It is interesting enough to see the
extra Higgs boson
also playing an important role in cancelling the Higgsino contribution to the $SU(2)_L\times U(1)_Y$ mixed anomaly, $U(1)_Y^3$ anomaly,
gravitational trace anomaly, and also to cancel the global $SU(2)_L$ Witten anomaly \cite{wittenanomaly}.

The SM or MSSM has been an extremely successful theory with exception of the puzzles, such as flavor hierarchy,
neutrino masses, the $\mu$-term problem, $R$-parity, the strong CP problem, etc.

In this thesis, our goal is to apply a new model building tool --- discrete gauge symmetries \cite{kw}
to solve the problems or puzzles mentioned above \cite{wk1,wk2,wk3,wk4}.

\section{Global Symmetries in the SM}

The SM provides one highly successful description the particle physics up to $M_{\rm E\rm W}$. However, it is believed to be only an
effective field theory valid up to a cutoff scale $\Lambda$. In the low-energy
effective theory, corrections from new physics beyond SM arise as
non-renormalizable operators which are invariant under $G_{\rm S\rm M}$. Unlike
the renormalizable couplings, the coupling constants of these
non-renormalizable operators are expected to be suppressed by appropriate powers of $1/{\mathrm
\Lambda}$ and have thus a negative dimension of mass \cite{operators,zee}.

In the SM, there is a unique unbroken $U(1)$ gauge symmetry which
is known as the $U(1)_Y$ hypercharge symmetry. The hypercharge
assignment except of its normalization is determined by requiring the
theory to be free from triangle gauge anomalies \cite{anomaly}.
The gauge anomalies are violations of conservative laws due to loop corrections.
They are generated via the triangle diagrams. For example, the $[SU(3)_C]^2\times U(1)$ mixed anomaly
arise from the following diagram:
\begin{center}
\begin{figure}[ht]
\begin{center}
  \includegraphics[scale=1,width=6cm, bb=0 0 300 250]{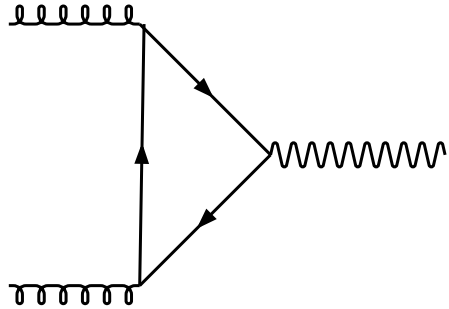}\end{center}
  \caption{The diagram that generates $[SU(3)_C]^2\times U(1)$ mixed anomaly.}
    \end{figure}
 \end{center}
where the internal lines are fermions, quarks in this case.

Being free from triangle gauge anomalies is a required condition for any gauge theory to make essential sense, namely  the renomalizability.
The anomalous Ward identity must be avoided. Anomaly matching condition should be satisfied.
In order to make the discussion more concrete, let us look at the explicit example of computing anomaly coefficients invoking the hypercharge
symmetry. Suppose under the $U(1)_Y$, $q$, $u$, $d$, $l$, $e$ and $h$ are corresponding charge for $Q$, $u^c$, $d^c$ $\ell$, $e^c$ and $H$.
$U(1)_Y$ invariance of the SM Yukawa couplings shown in \ref{smyukawa} requires \beq q+u+h=0,~q+d-h=0,~l+e-h=0.\eeq
The mixed anomaly coefficients should all vanish.
\bea
A_{[SU(3)_C]^2\times U(1)_Y}&=& \frac{N_g}{2}(2q+u+d)=0\nn\\
A_{[SU(2)_L]^2\times U(1)_Y}&=& \frac{N_g}{2}(3q+l)=0\nn\\
\rm{Tr} U(1)_Y&=& N_g(6q+3u+3d+2l+e)=0\nn\\
A_{[U(1)_Y]^3}&=& N_g(6q^3+3u^3+3d^3+2l^3+e^3)=0,
\eea
where the trace is the gravitational anomaly. In this particular case, cubic anomaly condition
is equivalent to the gravitational anomaly condition. One can then solve the set of 6 independent equations and obtain the hypercharge
assignment without its overall normalization. The hypercharge normalization can be determined when imposing conditions
from physics beyond SM, e.g., GUTs.

When the anomaly
cancellation constraints are relaxed, the extra degrees of freedom
correspond to the following global symmetries:
\begin{itemize}
\item~~ Baryon number $B$
\item~~ Lepton number $L$.
\end{itemize}They
cannot be realized as part of a fundamental gauge symmetry. An
ultimate theory, like string theory \cite{stringbook}, is believed to contain a theory of gravity which
presumably violates all global symmetries and therefore has to be a full gauge theory. It is then unclear where these global
symmetries arise from and how they can survive
down to low energies. One usually expects that global symmetries can
arise as accidental symmetries in the low energy effective theory.
However, there is  still no fundamental reason for global symmetries
to be protected. Both $B$ and $L$ could be violated but their violation
has not been directly observed yet.
When an additional Higgs doublet is introduced, the new degree of freedom correspond the Peccei-Quinn (PQ) symmetry \cite{PQ}.
The PQ symmetry is broken explicitly
near $f_a\sim {\mathcal O} (10^{11}{\mathrm G\rm e\rm V})$ thereby generating an
axion to compensate the CP violation in QCD and thus provides a
solution to the strong CP problem \footnote{ A detailed discussion can be found in chapter 6.}.

\begin{table}[ht]\label{global}
 \begin{center}
  {\renewcommand{\arraystretch}{1.1}
 \begin{tabular}{c c c c c c c c c}
   \hline
  \rule[5mm]{0mm}{0pt} & $Q$ & $u^c$ & $d^c$ & $\ell$ & $e^c$ & $\nu^{c}$& $H_u$ & $H_d$\\
  \hline
     \rule[5mm]{0mm}{0pt}
 $U(1)_B$& $1/3$&$-1/3$&$-1/3$&0&0&0&0&0\\
    \rule[5mm]{0mm}{0pt}
 $U(1)_L$& 0&0&0&$1$&$-1$&$-1$&0&0\\
    \rule[5mm]{0mm}{0pt}
     $PQ$ & 0 & 0 & $-1$ &
   0 & $-1$ & 0 & 1 & 0 \\
   \hline
 \end{tabular}
 }
  \caption{Global Symmetries in the two-Higgs SM}
 \end{center}
\end{table}
In Table \ref{global}, we list the global charges with respect to $U(1)_B$, $U(1)_L$ and the PQ symmetry for the two-Higgs SM which can be
naturally embedded into a SUSY version of the SM.

Since neither $B$ nor $L$ is a part of $G_{\rm S\rm M}$,
quantum gravity is believed to violate both $B$ or $L$ via non-renormalizable operators of the type: \beq
{\mathcal L}_{\rm N\rm R}\supset \ell\ell H H/M_{\rm P\rm l}+QQQ\ell/M^2_{\rm P\rm l}+\rm{h.c.}\eeq

The first term violates $L$ by two units ($\Delta L=2$)
and can give rise to neutrino masses, while the second term violates
both $B$ and $L$ by one unit ($\Delta B=1,~~\Delta L=1$) which leads to proton decay, for
example, via $p\rightarrow e^+ \pi^{0}$. Provided the four-dimensional (4D) quantum gravity
scale of $M_{\rm P\rm l}$ is roughly of order ${\mathcal O}(10^{19}~{\rm G\rm e\rm V})$,
one obtains a lower bound on the neutrino masses ($m_\nu$) and a upper bound on the proton lifetime ($\tau_p$) of the orders:
\beq
m_{\nu}\gtrsim 10^{-5}~ {\rm e\rm V}~~{\rm a\rm n\rm d}~~\tau_{p}\lesssim 10^{45} ~{\rm y\rm r\rm s}.\nn \eeq

The above neutrino mass scale does not agree with the current experimental
bound. For several decades, massless neutrinos have played an
important role in understanding the chiral character of weak
interaction. The SM does not contain massive neutrinos. However, since Super-Kamiokande
water Cherenkov Detector discovered the oscillation between different
flavor states of neutrinos suggesting that neutrinos are massive, our knowledge about neutrino masses has been remarkably improved by
solar \cite{soloar}, atmospheric \cite{atmospheric}, and reactor \cite{KamLAND} neutrino oscillation data. For
instance, solar and atmospheric neutrino oscillations imply the neutrino mass squared splittings $\Delta m^2_\odot =7.5\times 10^{-5}~ {\rm e\rm V}^2$ and
$\Delta m^2_\rm{atm} =2.0\times 10^{-3}~ {\rm e\rm V}^2$ respectively. These mass squared splittings yield a lower bound on
neutrino mass around $\sim 10^{-1}~ {\rm e\rm V}\gg 10^{-5}~ {\rm e\rm V}$ which is
much greater than the mass possibly induced by quantum gravity effects.

Lepton number does not necessarily have to be violated in order to understand the existence
of massive neutrinos. Neutrinos could be Dirac particles, in which case, neutrino masses may  arise from the usual Yukawa couplings:
\beq {\mathcal L}_\rm{Dirac}=\ell \nu^c H_u+\rm{h.c.}\eeq where right-handed neutrinos $\nu^c$ are the SM singlet. Then, the hierarchy problem in
Yukawa coupling constants must be addressed since there exists a
$10^{12}$ order hierarchy in $Y_{t}/Y_{\nu}\sim m_{t}/m_\nu\sim 174~{\rm G\rm e\rm V}/10^{-10}~{\rm G\rm e\rm V}\sim 10^{12}$.
The hierarchy provides a strong hint that a new physics scale should be much greater than $M_{\rm E\rm W}$. One natural way to understand this hierarchy,
i.e., the smallness of neutrino masses, is provided by the seesaw mechanism \cite{seesaw}.
In this framework, the right-handed neutrinos are Majorana particles and the right-handed neutrino scale is
$M_R \sim 10^{14}-10^{15}~{\rm G\rm e\rm V}$. The renormalizable lagrangian responsible for neutrino masses is then given by
\beq {\mathcal L}_{\rm s\rm e\rm e\rm s\rm a\rm w}=\ell \nu^c H_u+M_R \nu^c
\nu^c+{\rm h.c.}\eeq   Note that the Majorana
neutrino mass terms $M_R\nu^c\nu^c$ explicitly break the $L$. This allows to test the scenario in current and future neutrinoless
double beta  decay $(\beta\beta)_{0\nu}$ experiments. At low energies, the non-renormalizable $L$-violating operators generated by the seesaw mechanism
can be realized after integrating out the heavy right-handed neutrino as the dimension-five term\cite{zee,leung}
\beq {\mathcal L}_{\Delta L=2}=\ell\ell H H/{\Lambda_L},\eeq
where $\Lambda_L$ stands for the effective scale of $L$-violation which is $M_{R}$ in this case. After integrating out the heavy states $\nu^c$, one arrives at realistic neutrino
masses in the range  $M^2_{\rm E\rm W}/M_{R}\sim 10^{-10}~{\rm G\rm e\rm V}$.

The $\tau_p\sim 10^{45}~{\rm y\rm r\rm s}$ limit predicted by quantum gravity
corrections from operators of the type $QQQ\ell/{M^2_{\rm P\rm l}}$ is much above the current
experimental bounds on the proton lifetime \cite{pdg}:
\beq \tau_p >5\times 10^{33}~\mathrm{yrs}~~{\rm f\rm o\rm r}~~p\rightarrow e^+ \pi^{0}~~{\rm a\rm n\rm d}~~
 \tau_p >1.6\times 10^{33}~\mathrm{yrs}~~{\rm f\rm o\rm r}~~p\rightarrow \bar{\nu}K^{+}.\nn\eeq
These limits indicate that the baryon number violation scale must be $\Lambda_B>10^{15}~\mathrm{GeV}$.

The high energy scales $\Lambda_L$ and $\Lambda_B$ find a natural origin in Grand Unified Theories (GUTs) \cite{gut}. As
an elegant extensions of the SM, GUTs provide
a unified picture of the SM gauge interactions $SU(3)_C\times SU(2)_L\times U(1)_Y$
and are consistent with the gauge unification picture which LEP and other experiments
tested many years ago \cite{luo}. GUTs give a natural explanation of charge quantization
as well. As a result of putting baryons and leptons in to the same gauge multiplets, GUTs (with or without
SUSY) typically generate $\Delta B=1$ and $\Delta L=1$ operators with
\bed \Lambda_B\sim 10^{14}-10^{16}~\mathrm{GeV},\eed which are close to
the experimental limits from nucleon decay \cite{GQW}.

Besides the new physics effects discussed above,
$B$ can be violated even in the SM via non-perturbative effect such as electroweak instanton \cite{thooft}
and sphaleron processes \cite{ross}. These effects, however, are at the $\Delta B=3 ~{\rm m\rm o\rm d}~ 3$ level due to
the existence of three generations. For instance, the non-perturbative sphaleron interaction in the SM lagrangian can be
thought of as state \beq \prod_{i=1}^{3}(u_L d_L d_L \nu_L)_{i},\eeq where $i=1, 2, 3$ stands for generation index.
These $B$ and $L$ violating processes play an extremely important role in cosmology, e.g., in the context of  baryogenesis or the electroweak
phase transition. It is interesting to note that there exists a symmetry known as baryon parity \cite{ibanez1,wk3} in the SM lagrangian.
The physical consequence of this symmetry is also an effective Baryon number at the mod three level ($\Delta B=3~ {\rm m\rm o\rm d}~ 3$).
In the next chapter, we present this symmetry and discuss its physical implications.

\chapter{\textsc{Hidden Symmetry in the SM}}

\section{Discrete Gauge Symmetry and Anomalies}

Discrete global symmetries have been widely discussed in particle
physics for various phenomenological purposes. As mentioned previously,
global symmetries will have to face a potential violation induced by
quantum gravitational effects \cite{hawking}. If those discrete symmetries can be realized as gauge symmetries, such violation can then be avoided.
The idea of discrete gauge symmetries was first introduced in the Lattice gauge theory \cite{wegner}. One can make use of these discrete gauge symmetries
for field regularization purpose on the lattices \cite{lattice}. In the context of
string theory, discrete gauge symmetries are also widely discussed
as relics, emerging after dimensional reduction, of higher-dimensional general coordinate invariance or
spontaneously broken high-dimensional gauge symmetries. Moreover, they turn out to be
crucial in orbifold constructions \cite{gswbook}.
Discrete gauge symmetries are also introduced in 4D
field theories as remnants of a spontaneously broken gauge symmetry \cite{kw,ptww,banksdine}. As a new model building tool, discrete gauge symmetries have been widely discussed in various
applications \cite{ibanez1,ibanez2,ibanez3,yanagida,wk1,wk2,wk3,wk4,dias,banksdine2}.

In order to understand the idea of discrete gauge symmetries, let us consider an explicit realization of
a discrete gauge symmetry in a $U(1)$ theory.
Assume a 4D $U(1)$ gauge theory containing two scalars fields,
the Higgs $\eta$ with charge $N$ and the scalar $\psi$ with charge $-1$ under the $U(1)$ symmetry. After the Higgs
$\eta$ develops a vacuum expectation value (VEV) and breaks the $U(1)$, the gauge-invariant term $\eta \psi^N$ restricts
\beq \psi\rightarrow e^{-i 2\pi/N} \psi. \eeq However, since
the term $\eta \psi^N$ is non-renormalizable if $N\geq 4$, it is not clear whether the symmetry should really be
preserved. A renormalizable example in a chiral theory can be given in terms of the SM language, where masses arise from usual Yukawa
couplings. For this purpose, we suppose there exists a new $U(1)_X$ symmetry. Thus the total gauge symmetry of
the theory is $G_{\rm S\rm M}\times U(1)_X$. Suppose $U(1)_X$ is broken along the electroweak symmetry via
the SM Higgs VEV. The Yukawa coupling invariance then leads to
\beq Q u^c H: q+u+h=0 \eeq where $q$, $u$ and $h$ stand for the $U(1)_X$ charges of $Q$, $u^c$, and $H$, respectively. Hence, the fields transform as
\beq
 Q \rightarrow  e^{-i q \theta(x)} Q,~ u^c \rightarrow e^{-i u \theta(x)} u^c,~ H \rightarrow  e^{-i N \theta(x)} H,
\eeq where we also assume all the charges are integers and have set $h=N$.
After electroweak symmetry breaking, the lagrangian exhibits a  discrete $Z_N$ symmetry
\beq Q u^c \rightarrow e^{-i (q+u) 2\pi/N} Q u^c= e^{-i
(N) 2\pi/N}\ Q u^c =Q u^c,\eeq under which $Q$ and $u^c$ transform as \beq
 Q \rightarrow  e^{-i q 2\pi/N} Q,~
 u^c \rightarrow  e^{-i u^c 2\pi/N} u^c.
\eeq

In the effective theory, the two discrete $Z_N$ symmetries are
indistinguishable. However, this indeed provides hints to high
energy theory. Our above consideration provides a constraint on the proper charge assignment.
In fact, a condition must be satisfied, since spontaneous symmetry breaking
does not induce any gauge anomaly. Therefore, if the $Z_N$ is a subgroup of a gauge symmetry,
it must be free of gauge anomaly since the original theory is also anomaly-free.

Another puzzle arises as how to define a gauge anomaly in terms of discrete gauge symmetries \cite{kw,ptww,banksdine}. At low energies,
gauge bosons decouple from the theory and there is no gauge current associated with discrete gauge symmetries.
 It seems then to be difficult to realize a triangle anomaly \cite{anomaly}. However, as we mentioned earlier, gauge anomalies cannot be induced
 via spontaneous symmetry breaking (SSB), it should be possible to realize the anomaly prior to SSB. We can simply take the
 discrete charges to compute anomaly coefficients in the same way we compute anomalies before SSB. It is clear that
 the linear conditions will still hold. However, the non-linear conditions like cubic anomalies cannot be simply extended to discrete symmetries.

Besides the above change, the anomaly cancellation condition may be modified due to possible existence of vectorial heavy fermions. Suppose the discrete
$Z_N$ gauge symmetry arise from a full $U(1)$. The field that acquires a VEV and breaks $U(1)$ to $Z_N$ can
supply large masses at very high scale to a set of heavy fermions
which have Yukawa couplings involving this field. Such fields may
include Majorana fermions as \beq {\mathcal L}\supset S Q Q, \eeq and Dirac fermions as \beq {\mathcal L}\supset S Q\bar{Q}. \eeq These heavy fields
can carry SM gauge quantum numbers, but they must transform
vectorially under the SM. In order that their mass terms be
invariant under the unbroken $Z_N$, it must be that
\begin{eqnarray}
2 q_i &=& 0 ~mod~ N ~(\rm{Majorana ~fermion}) \nonumber \\
q_i + \bar{q_i} &=& 0 ~mod~N ~(\rm{Dirac ~fermion})
\end{eqnarray}
where $q_i$ are the $U(1)$ charges of these heavy fermions.  The
index $i$ is a flavor index corresponding to different heavy
fields. These heavy fermions, being chiral under the $U(1)_A$,
contribute to gauge anomalies.  Their contribution to the
$[SU(3)_C]^2 \times U(1)$ gauge anomaly is given by $A_3 =
\sum_i q_i m_i = (N/2)\sum_ip_i m_i$ (Majorana fermion) or
$A_3 = \sum_i (q_i + \bar{q_i})m_i = (N)\sum_ip_i m_i$ (Dirac
fermion) where $\ell_i$ is the quadratic index of the relevant
fermion under $SU(3)_C$ and the $p_i$ are integers. We shall adopt
the usual normalization of $m = 1/2$ for the fundamental of
$SU(N)$.  Then, for the case of a heavy Dirac fermion, one has $A_3
= p (N/2)$ where $p$ is an integer, as the index of the lowest
dimensional (fundamental) representations is 1/2 and those of all
other representations are integer multiples of 1/2. The same
conclusion follows for the case of Majorana fermions for a
slightly different reason.  All real representations of $SU(3)_C$
(such as an octet) have integer values of $m$, so that
$\sum_ip_i m_i$ is an integer.  Analogous conclusions follow
for the $[SU(2)_L]^2 \times U(1)$ anomaly coefficient.

\section{Baryon Parity}

In this section, we show that the SM lagrangian with the seesaw mechanism for small neutrino masses has a discrete
$Z_6$ gauge symmetry which forbids all $\Delta B=1$ and $\Delta
B=2$ baryon violating effective operators \footnote{Since there exists an unbroken $U(1)_Y$ symmetry, one can always take the hypercharge subgroup to
redefine the discrete symmetry as \bed H\rightarrow e^{-i 2\pi/3 \times(1)}
\cdot e^{-i 2\pi \alpha/N \times(3)}   H.\eed For instance, under all the symmetries we discuss here, Higgs fields transform non-trivially which
may lead to potential domain wall problem. But one can always rotate it away by shifting a combination of hypercharge.
We would then instead obtain a $Z_9$ symmetry. }. This can be seen as
follows. The SM Yukawa couplings incorporating the seesaw
mechanism to generate small neutrino masses is \beq {\mathcal
L}_\rm{mass} = Qu^{c}H+Qd^{c}{\bar{H}}+\ell
e^{c}{\bar{H}}+\ell\nu^{c}H+M_R \nu^{c}\nu^{c}+\rm{h.\rm c.}\eeq Here we have
used the standard (left-handed) notation for the fermion fields and
have not displayed the Yukawa couplings or the generation indices.
This lagrangian respects a  discrete $Z_6$ symmetry with the charge
assignment as shown in Table \ref{z6c}. Also shown in Table
\ref{z6c} are the charge assignments under the $Z_3$ and
$Z_2$ subgroups of $Z_6$. The $Z_3$ assignment is identical to
that in Ref. \cite{ibanez}
\begin{table}[ht]\label{z6c}
 \begin{center}
  {\renewcommand{\arraystretch}{1.1}
 \begin{tabular}{c c c c c c c c  }
   \hline
  \rule[5mm]{0mm}{0pt} & $Q$ & $u^c$ & $d^c$ & $\ell$ & $e^c$ & $\nu^{c}$& $H$ \\
  \hline
  \rule[5mm]{0mm}{0pt}
 $Z_6$&6 & 5 & 1 & 2 & 5 &3 & 1  \\
 $Z_3$&3 & 2 & 1 & 2 & 2 &3 & 1  \\
 $Z_2$&2 & 1 & 1 & 2 & 1 &1 & 1  \\
     \hline
\end{tabular}
 }
  \caption{Family-independent $Z_6$ charge assignment to the SM fields
   along with the charges under the $Z_3$ and $Z_2$ subgroups.  }
 \end{center}
\end{table}

From Table \ref{z6charge} it is easy to calculate the $Z_6$
crossed anomaly coefficients with the SM gauge groups. We find the
$SU(3)_C$ or $SU(2)_L$ anomalies to be\begin{eqnarray}
A_{{[SU(3)_C]}^2\times Z_6}=3N_g\nonumber\\
A_{{[SU(2)_L]}^2\times Z_6}=N_g~
\end{eqnarray}where $N_g$ is the number of generations.
The condition for a $Z_N$ discrete group to be anomaly-free is
\beq A_i=\frac{N}{2}~~\rm{mod}~N~\eeq where $i$ stands for $SU(3)_C$
and $SU(2)_L$. For $Z_6$, this condition reduces to $
A_i=3~\rm{mod}~6$, so when $N_g=3$, $Z_6$ is anomaly-free. Obviously,
the $Z_3$ and $Z_2$ subgroups are also anomaly-free. The
significance of this result is that unknown quantum gravitational
effects will respect this $Z_6$. It is this feature that we
utilize to stabilize the nucleon. Absence of anomalies also
suggests that the $Z_6$ may have a simple gauge origin.

To see how the $Z_6$ forbids $\Delta B=1$ and $\Delta B=2$
processes, we note that it is a subgroup of $U(1)_{2Y-B+3L}$ where
$Y$ is SM hypercharge \cite{wk}. We list in Table \ref{charge} the
charges under the three $U(1)$ symmetries.
\begin{table}[ht]
 \begin{center}
  {\renewcommand{\arraystretch}{1.1}
 \begin{tabular}{c c c c c c c c  }
   \hline
  \rule[5mm]{0mm}{0pt} & $Q$ & $u^c$ & $d^c$ & $\ell$ & $e^c$ & $\nu^{c}$& $H$ \\
  \hline
  \rule[5mm]{0mm}{0pt}
  $U(1)_{2Y-B+3L}$&0&$-1$&1&2&$-1$&$-3$&1\\
      \hline
\end{tabular}
 }
  \caption{ Charge assignment under $U(1)_{2Y-B+3L}$ which contains the $Z_6$.}
  \label{charge}
 \end{center}
\end{table}
It is clear that the $Z_6$ can be a subgroup of
$U(1)_{2Y-B+3L}$. Any $Z_6$ invariant effective operator must then
satisfy \beq \label{zz} 2\Delta Y-\Delta B+3\Delta L=0~~\rm{mod}~6.\eeq
Invariance under $U(1)_Y$ implies $\Delta Y=0$. Consider $\Delta
B=1$ effective operators which must then obey (from Eq.
(\ref{zz})) $3\Delta L=1~~\rm{mod}~6$. This has no solution, since
$3\Delta L=0~~\rm{mod}~3$ from Table \ref{charge}. Similarly, $\Delta
B=2$ operators must obey $3\Delta L=2~~\rm{mod}~6$ which also has no
solution. $\Delta B=3$ operators, which corresponds to $3\Delta
L=0~~\rm{mod}~6$, are allowed by this $Z_6$. Such operators have
dimension 15 or higher and have suppression factors of at least
$\Lambda^{-11}$. These will lead to ``triple nucleon decay"
processes where three nucleons in a heavy nucleus undergo
collective decays leading to processes such as $pnn\rightarrow
e^+\pi^0$. We estimate the rates for such decay in Section 2.3 and
find that $\Lambda$ can be as low as $10^2~\mathrm{GeV}$.

\section{Triple Nucleon Decays}

The existence of baryon parity ensures the absence of $\Delta B=1$ and $\Delta B=2$ effective operators.
We now list the lowest dimensional (d=15) $\Delta B=3$ effective
operators which are consistent with the baryon parity. Imposing
gauge invariance and Lorentz invariance, we find them to be:
\begin{eqnarray}\label{op} &&~~\bar{u^c}^{4}\bar{d^c}^{5}\bar{e^c},
~~{\bar{u^c}}^{2}{\bar{d^c}}^{7}{e^c},
~~Q{\bar{u^c}}^{3}{\bar{d^c}}^{5}\ell,
~~Q{\bar{u^c}}^{2}{\bar{d^c}}^{6}{\bar{\ell}},
~~Q^{2}{\bar{u^c}}^{3}{\bar{d^c}}^{4}{\bar{e^c}},\nonumber\\&&
~~Q^{2}{\bar{u^c}}{\bar{d^c}}^{6}{e^c},
~~Q^{3}{\bar{u^c}}^{2}{\bar{d^c}}^{4}\ell,
~~Q^{3}{\bar{u^c}}{\bar{d^c}}^{5}{\bar{\ell}},
~~Q^{4}{\bar{u^c}}^{2}{\bar{d^c}}^{3}{\bar{e^c}},
~~Q^{4}{\bar{u^c}}{\bar{d^c}}^{4}{\nu^c},\nonumber\\&&
~~Q^{4}{\bar{d^c}}^{5}{e^c},~~Q^{5}{\bar{u^c}}{\bar{d^c}}^{3}\ell,
~~Q^{5}{\bar{d^c}}^{4}{\bar{\ell}},
~~Q^{6}{\bar{u^c}}{\bar{d^c}}^{2}{\bar{e^c}}, ~~
Q^{7}{\bar{d^c}}^{2}\ell, ~~Q^{8}{\bar{d^c}}{\bar{e^c}}~.
\end{eqnarray} Here Lorentz, gauge and flavor indices are suppressed.
These operators can lead to ``triple nucleon decay". The dominant
processes are
\begin{eqnarray}\label{decay}
ppp&\rightarrow& e^{+}+\pi^{+}+\pi^{+}\nonumber\\
ppn&\rightarrow& e^{+}+\pi^{+}\nonumber\\
pnn&\rightarrow& e^{+}+\pi^{0}\nonumber\\
nnn&\rightarrow& \bar{\nu}+\pi^{0}~.
\end{eqnarray}

Tritium ($^{3}H$) and Helium-3 ($^{3}He$) are examples of
three-nucleon systems in nature. These nuclei are unstable and
undergo $\beta$-decay with relatively short lifetime. In the
presence of operators of Eq. (\ref{op}), $^{3}H\rightarrow
e^{+}+\pi^{0}$ and $^3 He\rightarrow e^{+}+\pi^{+}$ decays can
occur. However, there is no stringent experimental limit arising
from these nuclei. So we focus on triple-nucleon decay in the Oxygen
nucleus where there are experimental constraints from water
detectors. To estimate the decay lifetime we need  first to convert
the nine-quark operators of Eq. (\ref{op}) into three-nucleon
operators and subsequently into the Oxygen nucleus.

We choose a specific operator
$Q^5\bar{d^c}^4\bar{\ell}/\Lambda^{11}$ as an example to study the
process $pnn\rightarrow e^{+}+\pi^{0}$ triple nucleon decay
process. This induces the effective three-nucleon operator in the
Oxygen nucleus \beq\label{aa}
\frac{Q^5\bar{d^c}^4\bar{\ell}}{{\Lambda}^{11}}\sim\frac{\beta^3
(1+D+F)}{\sqrt{2}f_{\pi}{\Lambda^{11}}}(\pi nnpe)~,\eeq where
$\beta\simeq 0.014~{GeV}^{3}$ is the matrix element to convert
three quarks into a nucleon \cite{jlqcd}. $F\simeq 0.47$, $D\simeq
0.80$ are chiral lagrangian factors, and  $f_{\pi}=139$ MeV is the
pion decay constant.

We now estimate the wave-function overlap factor for three
nucleons inside Oxygen nucleus to find each other. This is based
on a crude free Fermi gas model where the nucleons are treated as
free particles inside an infinite potential well. A single nucleon
wave function is given by $\psi_{m}(x)=\sqrt{2/{r}}\sin({m\pi
x/{r}})$, where $r$ is the size of the nucleus and $m$ is the energy
level. Incorporating isospin and Pauli exclusion principle, the
highest energy level which corresponds to $m=4$ is found to have 2
protons and 2 neutrons. We assume the highest level has the highest
probability to form a Tritium-like ``bound state" of three
nucleons. The probability for three nucleons in the Oxygen nucleus to
overlap in a range of the size of the Tritium nucleus is \beq
P\sim\frac{4}{3}\int^{\sqrt[3]{\frac{3}{16}}}_{0}d\left({\frac{x_1}{r}}\right)d\left({\frac{x_2}{r}}\right)d\left({\frac{x_3}{r}}\right){\left(\sin\left({\frac{4\pi
x_1}{r}}\right)\sin\left({\frac{4\pi
x_2}{r}}\right)\sin\left({\frac{4\pi x_3}{r}}\right)\right)}^2\sim
0.0253~,\eeq where $\sqrt[3]{\frac{3}{16}}$ is the ratio between
the radii of the Tritium and the Oxygen nucleus, since $R \propto
A^{1/3}$ (A is the atomic number). So the effective baryon number
violating operator of Eq. (\ref{aa}) becomes \beq
\frac{P\beta^{3}}{\sqrt{2}f_{\pi}{\Lambda^{11}}R^{3}}( {^{3}H}\pi
e )~.\eeq The triple nucleon decay lifetime can then be estimated
to be \beq \tau\sim \frac{16\pi{f_{\pi}}^2{\Lambda}^{22}R^{6}}{
P^{2}\beta^{6} M_{^{3}H}~}.\eeq By putting the current limit on
proton lifetime of $3\times 10^{33}~\mathrm{yrs}$, we obtain:\beq
\Lambda\sim 10^2~\mathrm{GeV}~.\eeq Thus we see the $Z_6$ symmetry
ensures the stability of the nucleon. To test our crude model of nuclear transition, we have also
evaluated the double nucleon decay rate within the same approach
and found our results to be consistent with other more detailed
evaluations \cite{double}.

\section{Gauging Baryon Parity}

It is interesting to see if the $Z_6$ symmetry of Table
\ref{z6charge} can be realized as an unbroken subgroup of a gauged
$U(1)$ symmetry. Although the $Z_6$ is a subgroup of the
$U(1)_{(2Y-B+3L)}$, this $U(1)$ would be anomalous without
enlarging the particle content. We have found a simple and
economic embedding of $Z_6$ into a $U(1)$ gauge symmetry
associated with $I^{3}_{R}+L_i+L_j-2L_k$. Here $L_i$ is the $i$th
family lepton number and $i\neq j\neq k$. No new particles are
needed to cancel gauge anomalies. With the inclusion of
right-handed neutrinos, $I^{3}_{R}=Y-(B-L)/2$ is an anomaly-free
symmetry. $L_i+L_j-2L_k$, which corresponds to the $\lambda_8$
generator acting in the leptonic $SU(3)$ family space, is also
anomaly-free.

The charges of the SM particles under this $U(1)$ are
\begin{displaymath}
\begin{array}{cccc}
 Q_{i}=(0,0,0),& {u_{i}}^c=(-1, -1, -1), & {d_{i}}^c=(1, 1,
1),& \\
 \ell_{i}=(-4,2, 2),& {e_{i}}^c=(5, -1, -1),& {\nu_{i}}^c=(3, -3,
-3)~, &H=1.\\
\end{array}%
\nonumber
\end{displaymath}This
charge assignment allows all quark masses and mixings as well as
charged lepton masses. When the $U(1)$ symmetry breaks
spontaneously down to $Z_6$ by the vacuum expectation value of a
SM singlet scalar field $\phi$ with a charge of 6, realistic
neutrino masses and mixings are also induced. The relevant
lagrangian for the right-handed neutrino Majorana masses is
\beq\label{rh}
 {\mathcal L}_\rm{neutrino mass}=M_{12}\nu^c_1\nu^c_2+M_{13}\nu^c_1\nu^c_3+\nu^c_3\nu^c_3\phi+\nu^c_2\nu^c_2\phi+\nu^c_1\nu^c_1{\phi}^*
 +\nu^c_3\nu^c_2\phi~.
 \eeq
After integrating out the heavy right-handed neutrinos we obtain
the following $\Delta L=2$ effective operators: \beq
\label{bb}{\mathcal L}_{\Delta L=2}=\frac{1}{\Lambda}(\ell_1
\ell_2 H H+\ell_1 \ell_3 H H+\ell_1 \ell_1 H H \epsilon+\ell_2
\ell_2 H H{\epsilon}^{*}+\ell_2 \ell_3 H H{\epsilon}^{*}+\ell_3
\ell_3 H H{\epsilon}^{*}).\eeq Here $\Lambda\sim M_{12}\sim
M_{13}$ is the scale of $L$-violation and we have defined
$\epsilon\equiv\left<\phi\right>/\Lambda$. For $\epsilon\ll 1$,
this lagrangian leads to the inverted mass hierarchy pattern for
the neutrinos which is well consistent with the current neutrino
oscillation data. This neutrino mass mixing pattern is analogous
to the one obtained from $L_e-L_\mu-L_\tau$ symmetry \cite{babu}.
However, here the $U(1)$ is a true gauge symmetry.

We have also investigated other possible $U(1)$ origins of the
$Z_6$ symmetry and found the $I^3_R+L_i+L_j-2L_k$ combination to
be essentially unique. To see this, let us assign a general
$U(1)_X$ charge for $i$th generation of the SM fermions consistent
with the $Z_6$ symmetry as \bed
\{Q_i,u_i^c,d_i^c,\ell_i,e_i^c,\nu_i^{c}\}
=\{6m^{(i)}_1,5+6m^{(i)}_4,1+6m^{(i)}_3,2+6m^{(i)}_2,5+6m^{(i)}_5,3+6m^{(i)}_6\}\eed
where $m^{(i)}_j$ are all integers. The Higgs field has a charge
$H=1+6m_0$. If we impose the invariance of the Yukawa couplings of
the charged fermions and Dirac neutrinos for each generation, the
anomaly coefficients from the $i$th generation become
\begin{eqnarray}\label{anomaly}
A^{(i)}_{{[SU(3)_C]}^2\times U(1)_X}&=&0\nonumber\\
A^{(i)}_{{[SU(2)_L]}^2\times U(1)_X}&=& 1+9m^{(i)}_1+3m^{(i)}_2\nonumber\\
A^{(i)}_{{[U(1)_Y]}^2\times U(1)_X}&=& -(1+9m^{(i)}_1+3m^{(i)}_2)\nonumber\\
A^{(i)}_{{[U(1)_X]}^2\times U(1)_Y}&=& [5+m_0] A^{(i)}_{{[SU(2)_L]}^2\times U(1)_X}\nonumber\\
A^{(i)}_{{[U(1)_X]}^3}&=& [5+m_0]^{2}A^{(i)}_{{[SU(2)_L]}^2\times
U(1)_X}.
\end{eqnarray}
The coefficient for the mixed gravitational anomaly for each
generation is zero. From Eq. (\ref{anomaly}), it follows that
$A_2=\sum_i A^{(i)}_{{[SU(2)_L]}^2\times U(1)_X}=
\sum_i(1+9m^{(i)}_1+3m^{(i)}_2)=0$ can be satisfied only when all
three generation contributions are included. Once $A_2=0$ is
satisfied, all other anomaly coefficients will automatically
vanish. $A_2=0$ can be rewritten in a familiar form as $3\sum_i
Q_i+\sum_i \ell_i=0$. Thus we see that any $U(1)$ symmetry
satisfying this condition and consistent with the $Z_6$ charge
assignment can be a possible source of $Z_6$. If the $Q_i$ are
different for different generations, quark mixings cannot be
generated without additional particles. By making a shift
proportional to hypercharge, we can set $Q_i=0$ for all $i$. Two
obvious solutions to $\sum_i \ell_i=0$ are $\ell_i=(1,1,-2)$ and
$\ell_i=(1,-1,0)$. The latter one does not reproduce the $Z_6$
charge assignment while the former one does, which is our solution
when $I^3_R$ is added to it.

A related $B-3L_{\tau}$ has been discussed in Ref. \cite{ma}. This
is the same as $B-L$ plus $L_{e}+L_{\mu}-2L_{\tau}$. In Ref
\cite{ma}, only one right-handed neutrino $\nu^c_{\tau}$ is
introduced so the seesaw mechanism applies only for one light
neutrino. The other two neutrinos receive small masses from
radiative corrections. In our model, since there are three
right-handed neutrinos, all the neutrino masses arise from the
conventional seesaw mechanism.

The 4D simple GUT will explicitly
break the $Z_3$ baryon parity as it predicts the $D=6$ operator which violates the Baryon number
at $\Delta B=1$. One can also see that the embedding of $Z_6$ into $U(1)$ of $I^3_R+L_i+L_j-2L_k$
is not a consistent picture of the simple GUTs. The baryon parity provides a strong hint to GUTs type
physics beyond SM. It is interesting that the anomaly-free fact of $Z_6$ is a result of existence of
three generations and consistent with Baryon number violation due to the electric instanton or SM
sphaleron  processes.

\chapter{\textsc{Gauged $R$-parity and $B-L$ Symmetry}}

\section{MSSM and Gauged $R$-parity}

The following couplings
\beq {\mathcal W}_{\cancel{R}}=u^c d^c d^c+Q \ell d^c +\ell \ell e^c+ \ell H_u, \eeq are $G_\rm{SM}$ gauge invariant but absent in the non-supersymmetric
SM, since they violate Lorentz invariance. However, when the theory is extended to MSSM,
this constraint no longer exists and the couplings will appear in the superpotential. These couplings essentially violate $B$ or $L$
at the renormalizable level, which are presumably global symmetries in the SM. The $L$ violating couplings can give rise to the
neutrino masses via one-loop effects but the $B$ violating terms
lead to a rapid proton decay. The strong experimental bound on the proton lifetime therefore requires these couplings to be sufficiently suppressed, such
that the MSSM can become an acceptable theory. For this purpose, one usually assumes a discrete global $Z_2$ symmetry, under which,
 the SM particles are taken
to be even while their superpartners are odd. This symmetry is known as $R$-parity. The assumption of $R$-parity has profound implications for
supersymmetric particle search at colliders as well as for cosmology. Due to $R$-parity, e.g., SUSY particles would be produced at collders only in pairs.
Moreover, $R$-parity implies that the lightest SUSY particle (LSP), for instance a neutralino in the mSUGRA scenario, will be stable.
This stable LSP is then a leading
candidate for cosmological cold dark matter.

Since the global $R$-parity is not part of the MSSM gauge symmetry, it is potentially violated by quantum gravitational effects.
These effects (associated with worm
holes, black holes, etc.) are believed to violate all global symmetries \cite{hawking}.
Gauge symmetries, however, are  protected from such violations. As noted in Chapter 2, when a gauge symmetry breaks
spontaneously, often a discrete subgroup is left intact.  Such discrete symmetries, called discrete gauge symmetries \cite{kw}, are also immune to quantum gravitational effects.  Not all
discrete symmetries can however be gauge symmetries.  For instance, since the original continuous gauge symmetry was free from anomalies, its unbroken discrete subgroup should be free from
discrete gauge anomalies \cite{trivedi,db}.  This imposes a non--trivial constraint on the surviving discrete symmetry and/or on the low energy particle content
\cite{kw,trivedi,db,ibanez1,ibanez2,maru,ma}.
It will be of great interest to see if $R$--parity of MSSM can be realized as a discrete gauge symmetry, so that one can
rest assured that it wont be subject to unknown quantum gravitational
violations \cite{martin, wk1}.

After a systematic analysis in \cite{wk1}, we can conclude the simplest exact $R$-parity is family-independent $Z_2$ subgroup of $U(1)$ $I^3_R$ gauge symmetry.
\begin{table}[h]
 \begin{center}
  {\renewcommand{\arraystretch}{1.1}
 \begin{tabular}{c c c c c c c c c c }
   \hline
  \rule[5mm]{0mm}{0pt} & $Q$ & $u^c$ & $d^c$ & $\ell$ & $e^c$ & $\nu^{c}$& $H_u$ & $H_d$ & $\alpha$ \\
  \hline
  \rule[5mm]{0mm}{0pt}
 $Z_2$&2 & 1 & 1 & 2 & 1 &1 & 1 &1 &2 \\
     \hline
\end{tabular}
 }
  \caption{ Gauged $Z_2$ $R$-parity.}
  \label{z6charge}
 \end{center}
\end{table}
Notice together with the $Z_3$ Baryon Parity, they form a discrete $Z_6$ $R$-parity \cite{wk1}.

At the $U(1)$ level, $I^3_R$ can be realized as linear combination of $U(1)_{B-L}$ and $U(1)_Y$. Therefore, one can always take a
hypercharge subgroup to do a redefinition of an exact $Z_2$ $R$-parity. Hence, the exact $R$-parity can always be realized in terms of
subgroups of $B-L$ \cite{martin}.

Proton decay violates both $B$ and $L$. However,
$B-L$ is still conserved. The $U(1)_{B-L}$ is known as  a global symmetry in the SM. By introducing
a right-handed neutrino for each generation, it becomes a full gauge symmetry
free of anomalies. An interesting example of a discrete gauge symmetry in the SM
with seesaw neutrino masses is the $Z_6$ subgroup of
$B-L$.  The introduction of the right--handed neutrino for
generating small neutrino masses makes $B-L$ a true gauge
symmetry. When the $\nu^c$ fields acquire super-large Majorana
masses, $U(1)_{B-L}$ breaks down to a discrete $Z_6$ subgroup.
It is worth mentioning that the $Z_6$ symmetry has a $Z_2$ and a
$Z_3$ subgroup as well.

\begin{table}[h]
 \begin{center}
  {\renewcommand{\arraystretch}{1.1}
  \begin{tabular}{ c  c  c  c  c  c  c  c  c    }
     \hline

     \rule[5mm]{0mm}{0pt}
   Field & $Q$ & $u^{c}$ &
   $d^{c}$ & $\ell$ &
   $e^{c}$ & $\nu^{c}$ &
   $H_{u}$ & $H_{d}$  \\
   \hline
   U(1)$_{B-L}$ & 1/3 & $-1/3$ & $-1/3$ &
   $-1$ & 1 & $1$ & 0 & 0 \\
   \rule[5mm]{0mm}{0pt}
   $Z_6$ & $1$ & 5 & 5 &
   3 & 3 & 3 & 0 & 0 \\
   \hline
  \end{tabular}
  }
  \caption{ The $B-L$ charges of the SM fields
  along with the unbroken $Z_6$ subgroup after the seesaw mechanism.}
  \label{b-l}
 \end{center}
\end{table}

\section{Gauged $B-L$ without $\nu_R$}
One physical consequence of the existence of the $Z_3$ Baryon Parity in the theory is
that the new physics cut-off scale can be lowered to TeV even without violating current proton decay
limits. If the threshold of new physics is as low as a
few TeV, the induced neutrino mass via $\ell\ell H_u H_u/\Lambda$ will be too large. Here we show
a mechanism by which such operators can be suppressed by making
use of a discrete $Z_N$ symmetry (with $N$ odd) surviving to low
energy. This $Z_N$ has a natural embedding in the $B-L$ gauge symmetry.
The question arises here is essentially how to gauge $U(1)_{B-L}$ without a right-handed neutrino.

Consider the following effective operators in the low energy
lagrangian: \beq\label{cc} {\mathcal L}\supset\ell\ell H
H\frac{S^{6}}{\Lambda^{7}}+\frac{S^{2N}}{{ \Lambda^{2N-4}}}~.\eeq
Here $S$ is a scalar singlet field which has charge $(1,~3)$ under
$Z_N\times Z_6$ while $\ell$ has charge $(-3,~2)$.
The first term in Eq. (\ref{cc}) respects a $U(1)$ symmetry while
the second term reduces this to $Z_6\times Z_N$. If $S$ develops a
VEV of order $10^2~\mathrm{GeV}$, realistic neutrino masses can
arise even when $\Lambda$ is low. For example, if
$\Lambda=10~\mathrm{TeV}$ and $S=10^2~\mathrm{GeV}$, the neutrino
mass is of order $v^2 \left<S\right>^6/{\Lambda^{7}}\sim
0.4~\mathrm{eV}$, which is consistent with the mass scale
suggested by the atmospheric neutrino oscillation data.

Two explicit examples of the $Z_N$ symmetry with $N=5$ and $7$ are
shown in Table \ref{nu}. These $Z_N$ symmetries are free from
gauge anomalies. In the $Z_5$ example, the crossed anomaly
coefficients for $SU(3)_C$ and $SU(2)_L$  are $5N_g$ and $5N_g/2$
respectively showing that $Z_5$ is indeed anomaly-free. For $Z_7$,
these coefficients are $7N_g$ and $7N_g/2$, so it is also
anomaly-free.
\begin{table}[ht]\label{nu}
 \begin{center}
  {\renewcommand{\arraystretch}{1.1}
  \begin{tabular}{|c| c  c  c  c c   cc |   }
     \hline
     \rule[5mm]{0mm}{0pt}
   Field & $Q$ & $u^{c}$ &
   $d^{c}$ & $\ell$ &
   $e^{c}$  &
   $H$  & $S$\\
   \hline
   $Z_5$ & 1 & 4 & 4 &
   2 & 3  & 0  & 1\\
   $Z_7$ & $1$ & 6 & 6 &
   4 & 3  & 0  & 1\\
   \hline
  \end{tabular}
  }
  \caption{ $Z_N$ charge assignment for $N=5$ and $7$.}
  \label{z7}
 \end{center}
\end{table}

It is interesting to ask if the $Z_N$ can be embedded into a
gauged $U(1)$ symmetry. A simple possibility we have found is  to
embed this $Z_N$ into the anomalous $U(1)_A$ symmetry of string
origin with the anomalies cancelled by the Green-Schwarz mechanism
\cite{gs}. Consider $U(1)_{B-L}$ without the right-handed
neutrinos but with the inclusion of vector-like fermions which
have the quantum numbers of ${\bf 5}(3)$ and ${\bar{\bf 5}}(2)$
under $SU(5)\times U(1)_A$. This $U(1)_A$ is anomaly-free by
virtue of the Green-Schwarz mechanism. When this $U(1)_A$ breaks
down to $Z_5$, the extra particles get heavy masses and are removed
from the low energy theory which is the $Z_6\times Z_5$ model.

Without the second term in Eq. (\ref{cc}), the phase of the $S$ field
will be massless upon spontaneous symmetry breaking. This Majoron
field \cite{majoron} would however acquire a mass from the second
term of Eq. (\ref{cc}). In the $Z_6\times Z_5$ model, the mass of
the Majoron is of order $\left<S\right>^7/\Lambda^6\sim
100~\mathrm{keV}$. In the $Z_6\times Z_7$ model, the Majoron mass
is of order $\left<S\right>^{11}/\Lambda^{10}\sim 10~\mathrm{eV}$.
Such a Majoron with a mass of either $100~\mathrm{keV}$ or
$10~\mathrm{eV}$ is fully consistent with constraints from early
universe cosmology \cite{babumj}. The interaction term
$\ell\ell H H S^6/{\mathrm \Lambda}^7$ induces the Majoron decay
$S\rightarrow \nu\nu$ with a Yukawa coupling
$Y_{S\rightarrow\nu\nu}=6m_\nu/\left<S\right>\sim 10^{-11}$. The
decay rate of the Majoron can be estimated to be\beq
\Gamma=\frac{Y^2_{S\rightarrow\nu\nu}m_S}{8\pi}\sim 10^{-23}
m_S~.\eeq This corresponds to a Majoron lifetime of $\tau\sim
10~\mathrm{sec}$ for the $Z_6\times Z_5$ model and $\tau\sim
10^5~\mathrm{sec}$ for the $Z_6\times Z_7$ model. Such a Majoron
can modify the big-bang nucleosynthesis processes. However the
modification is not significant since the Majoron will decouple
before the electro-weak phase transition. Its contribution to the
expansion rate is equivalent to that of $0.047\times 4/7\sim
0.027$ light neutrino species \cite{nucleonsythen}. This extra
contribution is well within observational uncertainties.

\chapter{\textsc{The $\mu$-problem: A Symmetry Approach}}

The $\mu$ problem has been an intriguing puzzle of MSSM. One of the main goals of
SUSY is to solve the gauge hierarchy problem while the
arising $\mu$-problem brings the hierarchy problem back to the
theory \cite{susyreview,muterm}. $\mu$ and $B$ are the Higgs mass parameters in the MSSM, where $\mu$ appears in the superpotential
\beq W_\rm{MSSM}\supset \mu H_u H_d \eeq and $B$ is in the soft
breaking sector \beq {\mathcal L}_\rm{MSSM~soft}\supset \mu B
\tilde{H_u} \tilde{H_d}~.\eeq The phase of $B$ is also a main source of
CP violation in MSSM, usually known as the SUSY CP problem which
 is strictly constrained by electric dipole moment (EDM)
 experiments. To allow electroweak symmetry breaking, $\mu$ has to be of order the
 soft SUSY breaking mass scale $M_\rm{SUSY}$ (
$\sim M_\rm{EW}$), while one would usually expect $\mu$ to be of order the Planck scale $M_\rm{Pl}$,
the cut-off scale in the 4D theory, since it is a priori not protected by any (gauge)
symmetry. $\mu$ cannot vanish either, otherwise there would be massless
charged fermions (charged Higgsino).

Generally, understanding of the $\mu$ and $B$ parameters is usually tied up with the SUSY
breaking mechanism \cite{susy,susyreview}. In some scenario, it is directly related to the generation of gaugino masses \cite{muterm,gm}.

\section{Peccei-Quinn Symmetry}
In order to
explain the $\mu$ and $B$ parameters from physics beyond MSSM, one must introduce a new symmetry to ensure the absence of
the bare $\mu$ term in the superpotential and it is reasonable to
assume this new symmetry to be flavor independent. However, the fact
that under the new symmetry, $H_u$ and $H_d$ are not vectorial,
\beq h_u+h_d \neq 2\alpha,\eeq where $h_u$ and $h_d$ are the corresponding charges of the MSSM
two Higgs doublets and $\alpha$ is the charge of gaugino. This eventually leads to a global PQ symmetry. Suppose the new symmetry is an Abelian symmetry $G$.
After imposing the Yukawa couplings condition, the mixed QCD
anomaly is given as
 \begin{eqnarray}
A_{[SU(3)]^2\times G}&=&3\alpha+\frac{3}{2}(2(q-\alpha)+(u-\alpha)+(d-\alpha))\nonumber\\
&=& 3\alpha-\frac{3}{2}(h_u+h_d),
\end{eqnarray} where $q$, $u$, $d$ $h_u$ and $h_d$ are the corresponding charge under $G$ for the SSM superfields $Q$, $u^c$, $d^c$, $H_u$ and $H_d$.
$\alpha$ stands for the gaugino charge.

It is clear that this additional symmetry which forbids the bare
$\mu$ term in the superpotential carries mixed QCD anomaly so it
can be identified as the PQ symmetry \cite{PQ}.

A naive
extension of the MSSM is to introduce a SM singlet $S$ with trilinear coupling $H_u H_d S$ \cite{nmssm}. The quartic-coupling can arise
radiatively. After the PQ symmetry is broken, $S$ develops a VEV of order of $M_\rm{EW}$. Hence,
$\mu$ arises via \beq \mu\sim\left<S\right>\sim M_\rm{EW}.\eeq
Global PQ symmetry is explicitly broken by its mixed QCD anomaly
giving rise to a pseudo-Goldstone particle known as the axion.
The axion mass and PQ symmetry breaking scale could both
lead to phenomenological inconsistency \footnote{Details of axion physics are discussed in
Chapter 6.}.

Based on the different ways to address this PQ symmetry problem,
the solutions to the $\mu$ problem can be classified into the
following categories:
\begin{itemize}
\item ~~ Explicit breaking of PQ symmetry \cite{nmssm} \item ~~ Gauging the PQ symmetry by
adding exotic quarks \cite{u1p} \item ~~ Addition of a discrete global R-symmetry \cite{rsymmetry}
\item ~~ Realization as a subgroup of Anomalous $U(1)_A$ gauge symmetry \cite{gm,wk1}
\item ~~ Supersymmetric invisible axion solution \cite{axionmu,wk2,wk4,shafi}
\end{itemize}
The first solution is realized as Next-to-Minimal model (NMSSM), which is defined by
\beq \mathcal{W}_\rm{NMSSM}\supset H_u H_d S +S^3 \eeq where $S^3$
 breaks explicitly the $U(1)_\rm{PQ}$. However, at the same
time, a new $Z_3$ discrete global symmetry has to be introduced, where
$S$ transform non-trivially under the discrete
$Z_3$ symmetry. As a consequence, a domain wall is
formed at the scale $M_\rm{SUSY}$ which is much lower than the inflation
scale. So this poses a serious cosmological problem.

Following the second approach, a string motivated $U(1)'$ gauge
symmetry has been proposed \cite{u1p}. The $\mu$-term solution here is quite
similar to the NMSSM but without discrete $Z_3$ symmetry and involves the superpotential terms \beq
\mathcal{W}_{U(1)'}=h H_u H_d S+\lambda  S_1 S_2 S_3.\eeq $S$ gets a VEV
near $M_\rm{SUSY}$ from soft SUSY breaking sector. It arises
from a string originated $E_6$ symmetry  with symmetry breaking pattern \begin{displaymath}
E_6\rightarrow SO(10)\times U(1)_\psi \rightarrow SU(5)\times
U(1)_\chi \times U(1)_\psi,
\end{displaymath} so it is a full gauge symmetry. But the theory has
many exotic matter particles decoupled near one TeV. The $U(1)'$ also predicts an extra $Z'$ boson which may contribute to precision electroweak tests.

In the following three sections, we will present here  the last two solutions.

\section{Giudice-Masiero Mechanism}

One attractive scenario which achieves a $\mu$-term solution in the
SUGRA mediated SUSY breaking mechanism is the Guidice--Masiero
mechanism \cite{gm} where a bare $\mu$-term in the superpotential
is forbidden by some symmetry, either discrete or continuous.
$\mu$ is induced in the lagrangian via a non-renormalizable term
\begin{equation}
{\mathcal L} = \int d^4 \theta {H_u H_d Z^* \over M_\rm{Pl}}
\end{equation}
where $Z$ is a spurion field which parameterizes SUSY
breaking via $\left \langle F_Z \right \rangle \neq 0$, with
$\left \langle F_Z \right \rangle/M_\rm{Pl} \sim M_\rm{SUSY} \sim 10^2$ GeV. For instance, the gaugino masses are generated from
 \beq {\mathcal L}_\rm{soft}\supset \int d^2 \theta
W_\alpha W^\alpha \frac{Z}{M_\rm{Pl}} \eeq

For this mechanism
to work, there must exist a symmetry that forbids a bare $\mu$
term in the superpotential. Such a symmetry cannot be a continuous
symmetry, consistent with the requirement of non--zero gaugino
masses, and therefore must be discrete. \footnote{Without the
$\mu$-term and the gaugino mass term, the MSSM Lagrangian has two
$U(1)$ symmetries, a PQ symmetry and a $U(1)_R$
symmetry. The $\mu$-term breaks the PQ symmetry and the
gaugino mass term breaks the $U(1)_R$ symmetry down to a discrete
subgroup.} It would be desirable to realize this as a discrete
gauge symmetry so that the symmetry will be protected even at
$M_\rm{Pl}$. However, as mentioned previously, one must avoid
the PQ symmetry problem and one of the ways is to gauge the PQ
symmetry through the Green-Schwarz (GS) mechanism and realize the discrete
symmetry as a subgroup of the anomalous $U(1)_A$ gauge symmetry \cite{gs}.

\section{Green-Schwarz Anomaly Cancellation Mechanism}

String theory, when compactified to 4D, generically contains an
``anomalous $U(1)_A$" gauge symmetry.  A subset of the gauge
anomalies in the axial vector $U(1)_A$ current can be cancelled
via the Green-Schwarz (GS) mechanism in the following way \cite{gs}.
In 4D, the Lagrangian for the gauge boson kinetic energy
contains the terms
\begin{equation}
{\mathcal L}_\rm{kinetic} = \varphi(x) \sum_ik_{i} F_i^2 +
i\eta (x) \sum_ik_i F_i \tilde{F_i},
\end{equation}
where $\varphi (x)$ denotes the string dilaton field and $\eta
(x)$ its axionic partner.  The sum $i$ runs over the different
gauge groups in the model, including $U(1)_A$.  $k_i$ are the
Kac--Moody levels for the different gauge groups, which must be
positive integers for the non--Abelian groups, but may be
non--integers for Abelian groups.  The GS mechanism makes use of
the transformation of the string axion field $\eta (x)$ under a
$U(1)_A$ gauge variation,
\beq \label{gsss}
V^\mu_A \rightarrow V^\mu_A +
\partial_\mu \theta (x),~
\eta (x) \rightarrow \eta (x) -\theta (x) \delta_\rm{GS}
\eeq
where $\delta_\rm{GS}$ is a constant.  If the anomaly coefficients
involving the $U(1)_A$ gauge boson and any other pair of gauge
bosons are in the ratio
\begin{equation}
{A_1 \over k_1} = {A_2 \over k_2} = {A_3 \over k_3}=....={A_\rm{gravity}\over 24}=
\delta_\rm{GS}~,
\end{equation}
these anomalies will be cancelled by gauge variations of the
$U(1)_A$ field arising from the second term of Eq. \ref{gsss}.
$\delta_\rm{GS}$ is known as Green-Schwarz constant which is defined in term of the mixed gravitational
anomaly. All other crossed anomaly coefficients should vanish, since they
cannot be removed by the shift in the string axion field.

Consider the case when the 4D gauge symmetry  just
below the string scale is $G_\rm{SM}
\times U(1)_A$.  Let $A_3$ and $A_2$ denote the anomalies
associated with $[SU(3)_C]^2 \times U(1)_A$ and $[SU(2)_L]^2
\times U(1)_A$ respectively.  Then if $A_3/k_3 = A_2/k_2 =
\delta_\rm{GS}$ is satisfied, from Eq. (4), it follows that these
mixed anomalies will be cancelled.  The anomaly in $[U(1)_Y^2]
\times U(1)_A$ can also be cancelled in a similar way if $A_1/k_1
= \delta_\rm{GS}$.  However, in practice, this last condition is less
useful, since $k_1$ is not constrained to be an integer as the
overall normalization of the hypercharge is arbitrary.  If the
full high energy theory is specified, there can be constraints on
$A_1$ as well. For example, if hypercharge is embedded into a
simple group such as $SU(5)$ or $SO(10)$, $k_1 = 5/3$ is fixed
since hypercharge is now quantized.  $A_1/k_1 = \delta_\rm{GS}$ will
provide a useful constraint in this case.  We shall remark on this
possibility in our discussions.  Note also that cross anomalies
such as $[SU(3)] \times [U(1)_A]^2$ are automatically zero in the
SM, since the trace of $SU(N)$ generators is zero.
Anomalies of the type $[U(1)_Y] \times [U(1)_A]^2$ also suffer
from the same arbitrariness from the Abelian levels $k_1$ and
$k_A$.  Finally, the $[U(1)_A]^3$ anomaly can be cancelled by the GS
mechanism, or by contributions from fields that are complete
singlets of the SM gauge group.

As discussed in Section 2.1, discrete version of anomaly cancellation
will need to be modified due to the possible existence of vectorial fermion pairs.
If the $Z_N$ symmetry that survives to low energies was part of
$U(1)_A$, the $Z_N$ charges of the fermions in the low energy
theory must satisfy a non--trivial condition:  the anomaly
coefficients $A_i$ for the full theory is given by $A_i$ from the
low energy sector plus an integer multiple of $N/2$.  These
anomalies should obey GS mechanism, leading to the discrete version of
the Green--Schwarz anomaly cancellation mechanism:
\begin{equation}\label{dgsm}
{A_3 + {p_1 N  \over 2} \over k_3} = {A_2 + {p_2 N \over 2} \over
k_2} = \delta_\rm{GS},
\end{equation}
with $p_1, ~p_2$ being integers.  Since $\delta_\rm{GS}$ is an
unknown constant (from the effective low energy point of view),
the discrete anomaly cancellation conditions  are less
stringent than those arising from conventional anomaly
cancellations.  If $\delta_\rm{GS} = 0$, the anomaly is
cancelled without assistance from the Green--Schwarz mechanism. We
shall not explicitly use the condition that $\delta_\rm{GS} \neq 0$,
so our solutions will contain those obtained by demanding
$\delta_\rm{GS} = 0$, viz., $A_3 = -p_1 (N/2)$, $A_2 =
-p_2 (N/2)$ with $p_1,~p_2$ being integers.

The anomalous $U(1)_A$ symmetry is expected to be broken just
below the string scale.  This occurs when the Fayet--Iliopoulos
term associated with the $U(1)_A$ symmetry is cancelled, so that
SUSY remains unbroken near the string scale, by shifting
the matter superfields that carry $U(1)_A$ charges \cite{dsw}.
Although the $U(1)_A$ symmetry is broken, a $Z_N$ subgroup of
$U(1)_A$ can remain intact.  Suppose that we choose a
normalization wherein the $U(1)_A$ charges of all fields are
integers.  (This can be done so long as all the charges are
relatively rational numbers.)  Suppose that the scalar field which
acquires a vacuum expectation value (VEV) and breaks the $U(1)_A$
symmetry has a charge $N$ under $U(1)_A$ in this normalization.  A
$Z_N$ subgroup is then left unbroken down to low energies.

In our analysis we shall not explicitly make use of the condition
$A_1/k_1 = A_2/k_2$, since, as mentioned earlier, the overall
normalization of hypercharge is arbitrary.  However, once a
solution to the various $Z_N$ charges is obtained, we can check
for the allowed values  $k_1$, and in particular, if $k_1 = 5/3$
is part of the allowed solutions. This will be an interesting case
for two reasons.  If hypercharge is embedded in a simple grand
unification group such as $SU(5)$, one would expect $k_1 = 5/3$.
Even without a GUT embedding $k_1 = 5/3$ is interesting. We recall
that unification of gauge couplings is a necessary phenomenon in
string theory.  Specifically, at tree level, the gauge couplings
of the different gauge groups are related to the string coupling
constant $g_\rm{st}$ which is determined by the VEV of the
dilaton field as \cite{ginsparg}
\begin{equation}
k_3 g_3^2 = k_2 g_2^2 = k_1 g_1^2 = 2 g_\rm{st}^2~
\end{equation}
where $k_i$ are the levels of the corresponding Kac--Moody
algebra.  In particular, if $k_1: k_2:k_3 = 5/3:1:1$, we would
have $\sin^2\theta_W = 3/8$ at the string scale, a scenario
identical to that of conventional  gauge coupling unification with
simple group such as $SU(5)$.  For these reasons, we shall pay
special attention to the case $k_1 = 5/3$.

\section{Discrete $Z_4$ Gauge Symmetry from $U(1)_A$}

As discussed in the above two sections, the symmetry which is consistent with the
Giudice-Masiero mechanism must be discrete but carries a mixed QCD anomaly \cite{gm}.
So the only realization of discrete gauge symmetries must arise from the anomalous $U(1)_A$ gauge symmetry.
We have done a systematic analysis in \cite {wk1}.

Here, one of examples of $Z_4$ subgroup of the anomalous
$U(1)_A$ is given in Table {\ref{z4p}}.

\begin{table}[h]\label{z4p}
 \begin{center}
  {\renewcommand{\arraystretch}{1.1}
 \begin{tabular}{ c c  c c c c c c c  }
   \hline
  \rule[5mm]{0mm}{0pt}
  $q$ & $u$ & $d$ & $l$ & $e$ & $n$ & $h$ & $\bar{h}$ & $\alpha$  \\
  \hline
  \rule[5mm]{0mm}{0pt}
    1 & 1 & 1 & 1 & 1 & 1 & 0 & 0 & 1 \\
        \hline
\end{tabular}
 }
  \caption{ $Z_4$ subgroup of the Anomalous $U(1)_A$}
 \end{center}
\end{table}
The mixed anomaly coefficients are
\beq A_3=3~\rm{mod}~4,~~ A_2=1~\rm{mod}4\eeq which satisfies the discrete version of Green-Schwarz anomaly
cancellation condition. The charge assignment shown in Table
\ref{z4p} is clearly compatible with grand unification. The
Kac--Moody level associated with hypercharge will be $k_1 = 5/3$
with a GUT embedding.  Gauge coupling unification is then
predicted, since $\sin^2\theta_W = 3/8$ near the string scale.
This is true even if there were no covering GUT symmetry. It also
acts as a exact $R$-parity. The anomalous $U(1)_A$ is broken to  \begin{displaymath} U(1)_A\rightarrow
Z_4 \rightarrow Z_2, \end{displaymath} where after the SUSY breaking,
$Z_4$ is broken into the $Z_2$ subgroup of the $I^3_R$ as in chapter 3.

\section{QCD Axion Solution to the $\mu$ Problem}

As mentioned previously, the PQ symmetry implies the presence of an axion. It is interesting to note that an axion is required to
solve the Strong CP problem. All the acceptable axion solutions
must be ``invisible". Here, we present a model making use of the
real QCD axion to address the $\mu$-term problem. This is a natural solution in terms of the PQ symmetry while the QCD axion is an elegant
solution to the Strong CP problem and at the same time, it
provides candidate for cold dark matter.

In all the above approaches, the $\mu$-term solutions eventually make
use of SUSY breaking and directly relate the $M_\rm{SUSY}$ to
$\mu$. Imposing a new physics scale $M_\rm{PQ}~(f_a=(10^{10}-10^{12})~\rm{GeV})$, the axion models
provide another approach to the $\mu$-term problem\cite{axionmu,wk2,wk4,shafi} by relating\beq
\mu\sim \frac{M^2_\rm{PQ}}{M_\rm{Pl}}.\eeq In the case
of the Dine-Fischler-Srednicki-Zhitnitskii (DFSZ) axion model \cite{DFSZ}, a $\mu$-term automatically arises after PQ
symmetry breaking.

The question is now how to naturally understand the origin of
$M_\rm{PQ}$ from a higher energy theory. It is interesting
that in the SUGRA mediated SUSY breaking models, one also has to
impose a new physics scale of order ${\mathcal O}(10^{11}~\rm{GeV})$. In these models, this intermediate scale can be generated
dynamically. Practically, this intermediate scale can then be
identified as $M_\rm{PQ}$. Here we propose a model involving
SUSY breaking \cite{babu}. Having made use of $M_\rm{SUSY}$,
this approach certainly requires that the SUSY breaking mediation
scale is greater than $M_\rm{PQ}$. A simple realization of
this idea is the SUGRA model. The superpotential of the model
contains \beq \mathcal{W}\supset\lambda_1 H_u H_d S^2/{M_\rm{Pl}}
+{\lambda_2 (S\tilde{S})^{2}/{M_\rm{Pl}}}+{S^{22}}/{M^{19}_\rm{Pl}}\eeq which is also consistent
with the $Z_{22}$ symmetry in the previous section. By minimizing
the leading-order potential including SUSY breaking effects, \beq
V=(\lambda_{2}C{(S\tilde{S})^2}/{M_\rm{Pl}}+h.c)+{m_S}^2|S|^2+{m_
{\tilde{S}}}^2{|\tilde{S}|}^2+4\lambda_2
{|S\tilde{S}|^2}(|S|^2+{|\tilde{S}|}^2)/{M_\rm{Pl}^2},\eeq
where $m_S$ and $m_{\tilde{S}}$ are soft breaking masses of order
$M_\rm{SUSY}$, one obtains \beq
f_a^2={C\pm\sqrt{C^2-12{m_{S}}^2}}M_\rm{Pl}/{12\lambda_2}.\eeq So \beq f_a \sim \sqrt{M_\rm{Pl}M_\rm{SUSY}}
 \sim 10^{11}~\mathrm{GeV}.\eeq Since
the $F$-component of the field $S$ obeys \beq F_S\sim M_\rm{PQ}
M_\rm{SUSY},\eeq the dominant contribution for the $B$
parameter which appears in the soft bilinear SUSY breaking term
\beq {\mathcal L_\rm{soft}}\supset B\mu H_u H_d\eeq arises
from the superpotential $H_u H_d S^2/M_\rm{Pl}$ as \beq
B\mu=\langle S\rangle \langle F_S \rangle /M_\rm{Pl}\sim
M^2_\rm{SUSY}.\eeq So it is difficult to distinguish it from
the usual MSSM via electroweak physics. However, as the axion can
be a cold dark matter candidate, one can still distinguish the
model in cosmology. In this model, the two PQ Higgs bosons have
masses of order $M_\rm{SUSY}$ but their mixings with the
doublet Higgs are highly suppressed. The orthogonal combination to
the axion acquires a mass of order $M_\rm{SUSY}$. The axino
and saxino masses are both around $M_\rm{SUSY}$. The axino
can mix with the Higgsino with a tiny mixing angle of order
$(M_\rm{SUSY}/M_\rm{Pl})^{1/2}\sim 10^{-7}$. Therefore,
the axino can decay to a bottom quark and a sbottom squark with a
lifetime \beq \tau\sim 10^{-11}~\mathrm{sec}.\eeq This is a
consistent picture with big-bang cosmology since the axino decays
occur earlier than the nucleosynthesis era.

\chapter{\textsc{Discrete Flavor Gauge Symmetry}}

The flavor hierarchy problem has been a very challenging problem in model building for many years \cite{mass}.
It mainly addresses the following two questions:
\begin{itemize}
\item~~ How is the apparent $10^{12}$ order hierarchy in $m_{t}/m_{\nu}$ generated?
\item~~ What is the origin of the observed mass ratios and mixing angles of the SM quarks and leptons?
\end{itemize}

As mentioned earlier in the discussion of $L$ violation, neutrino masses provides a hint for a new physics scale when they are understood
as emerging in the low-energy theory from operators such as $\ell\ell H_u H_u/{\mathrm \Lambda_L}$. In the following, we use the seesaw mechanism,
one natural approach as the realization of this new physics:
\beq {\mathcal W}_\rm{seesaw}=\ell \nu^c H_u+M_R \nu^c \nu^c \eeq
where there exists heavy Majarona neutrinos $\nu^c$ at $M_R\sim {\mathcal O}(10^{14}~\rm{ GeV})$ and the small neutrino masses are given by $M^2_\rm{ EW}/M_{R}\sim 10^{-10}~\rm{ GeV}$.

In the SM quark and charged lepton sectors, the mixing angles are small. However,
there has been a strong evidence for large neutrino mixing from recent solar, atmospheric and reactor neutrino
oscillation data. It is then a challenge to understand why there exists such a discrepancy in the mixing angles, especially in the context of GUTs
where quarks and leptons are unified in the same GUT multiplets.

In a 4D framework, flavor gauge symmetries have been a leading candidate solution to this problem. And even in the string theory
which has achieved family unification in extra spacetime dimensions, flavor gauge symmetries exist as well. Here, the symmetries
are usually broke by  the boundary conditions explicitly. Such orbifold models, however, correspond to special points in the moduli space of the
Calabi-Yau manifold at which there is an extra gauge symmetry that acts on the flavors. The more generic Calabi-Yau models can then be
considered as models in which the flavor gauge symmetries are spontaneously broken.

SUSY is a promising candidate for physics beyond SM. But when SUSY is introduced, a new flavor problem arises in the soft breaking
sector known as the SUSY flavor problem. In the SM, one can make use of the GIM mechanism \cite{gim} to suppress harmful flavor
changing neutral currents (FCNC's) by a suitable unitary transformation between mass and gauge eigenstates. Alternatively: it is, however,
not clear why soft sfermion masses sector and the fermion masses sector in SUSY models should transform similarly. The difference between the
 usual Yukawa and soft breaking sectors may thus lead to flavor violation, which is strictly constrained from $K-\bar{K}$ mixing
and lepton flavor violation measurement like $\mu\rightarrow e\gamma$. This issue depends on the understanding of the
SUSY breaking mechanism as well as the flavor gauge symmetries. A popular solution is to assume universality in the soft breaking sector. Then universal
structure will remain universal after the unitary transformation. For instance, gauge mediated SUSY breaking or string dilaton dominant SUSY breaking
both provide a universal soft sector. In the most widely discussed SUGRA type models, people usually assume the universality.
However, any flavor gauge symmetry
will bring a splitting between different generations back to the theory known as the $D$-term splitting problem.
A discrete flavor symmetry, on the other hand, would avoid this problem as there is no $D$-term associated with it. In the following sections, we
present an explicit example of discrete flavor gauge symmetry approach.

\section{Froggatt-Nielsen Mechanism and Anomalous $U(1)_A$ Realization}

The most straightforward example of a flavor gauge symmetry is the $U(1)_F$ symmetry  employed as in the Froggatt-Nielsen mechanism \cite{fn}.
Here, a SM singlet scalar which couples to SM matter Yukawa terms is introduced, which and transforms under the new $U(1)_F$ symmetry. The quarks and leptons
also carry different $U(1)_F$ flavor charges. Some new physics generates the non-renormalizable couplings terms consistent
with $G_\rm{SM}\times U(1)_F$ invariance.  In terms of MSSM, the superpotential is given as
\begin{eqnarray}\label{fnn}
{\mathcal W}&=&\frac{y_{ij}^u}{n^u_{ij}!} Q_i u_j^c H_u
\left(\frac{S}{\mathrm \Lambda_\rm{FN}}\right)^{n^u_{ij}}+
\frac{y_{ij}^d}{n^d_{ij}!}
Q_id_j^c H_d \left(\frac{S}{\mathrm \Lambda_\rm{FN}}\right)^{n^d_{ij}}\nn \\
&+& \frac{y_{ij}^e}{n^e_{ij}!} L_i e_j^c H_d
\left(\frac{S}{\mathrm \Lambda_\rm{FN}}\right)^{n^e_{ij}}
 + \frac{y_{ij}^\nu}{n^\nu_{ij}!} L_i \nu^c_j H_u \left(\frac{S}{\mathrm \Lambda_\rm{FN}}\right)^{n^\nu_{ij}}\nonumber\\
 &+&{M_{R}}_{ij} \nu^c_i \nu^c_j\left(\frac{S}{\mathrm \Lambda_\rm{FN}}\right)^{n^{\nu^c}_{ij}}+\mu H_u
 H_d\, ,
\end{eqnarray}
where $i,j=\{1,\,2,\,3\}$ are family indices, ${n^u_{ij}}$,
${n^d_{ij}}$, ${n^e_{ij}}$, ${n^{\nu}_{ij}}$ and $n^{\nu^c}_{ij}$
are positive integers fixed by the choice of $U(1)_F$ charge
assignment. The quantities $y^x_{ij}$, where $x=u,d,e,\nu$, are Yukawa coupling coefficients which
are all taken to be of order one. Here, $M_R$ is the right-handed neutrino mass scale.

When the SM singlet $S$ acquires a VEV, the $U(1)_F$ symmetry is spontaneously broken. Hierarchy and mixings thus arise as suppression of different
powers of $S/\Lambda_\rm{FN}$.

In the Froggatt-Nielsen mechanism, the usual parametrization of the fermion mass matrices requires $S/\Lambda_\rm{FN}(\epsilon)\sim 1/5$.
The anomalous $U(1)_A$ symmetry which we discuss in Section 4.3 \cite{gs, dsw} is a promising realization here. The anomalous $U(1)_A$ symmetry is
broken below the string scale $M_\rm{St}\sim \mathcal{O}(10^{17}\rm{GeV})$. Hence,
a natural realization of $\epsilon$ comes as \beq \epsilon\sim \left<S\right>/M_\rm{St}\sim 0.2\eeq

\section{A Lopsided Structure and Discrete Flavor Gauge Symmetry}

As mentioned earlier, it is a challenge to address flavor hierarchy problem in a GUT framework.

At low energy, the fermion masses are \cite{gasser}
\begin{eqnarray}\label{ckm}
m_u(1~\rm{GeV})= 5.11~\rm{MeV}, & m_c(m_c)= 1.27~\rm{GeV}, &  m_t(m_Z)= 174~\rm{GeV},\nn\\
m_d(1~\rm{GeV})= 8.9~\rm{MeV}, & m_s(1~\rm{GeV})= 130~\rm{MeV}, & m_b(m_b)= 4.25~\rm{GeV}.
\end{eqnarray}
The CKM mixing matrix elements are \beq\label{cmk2} |V_{us}|\sim 0.222,~|V_{ub}|\sim 0.0035,~|V_{cb}|\sim 0.04.\eeq

In \cite{wk1}, we proposed an $SU(5)$ GUT compatible model. An acceptable flavor texture which gives the correct pattern of
fermion masses and mixings as shown in \ref{ckm} and \ref{ckm2} is:
\begin{eqnarray}
\label{matr1}
 U_{ij}&=&\left(
\begin{array}{ccc}
  \epsilon^{6} & \epsilon^{5} & \epsilon^{3} \\
  \epsilon^{5} & \epsilon^{4} & \epsilon^{2} \\
  \epsilon^{3} & \epsilon^{2} & 1 \\
\end{array}
\right)H_{u}, ~~~~ D_{ij}=\left(
\begin{array}{ccc}
  \epsilon^{4} & \epsilon^{3} & \epsilon^{3} \\
  \epsilon^{3} & \epsilon^{2} & \epsilon^{2} \\
  \epsilon & 1 & 1 \\
\end{array}
\right)\epsilon^{p}H_{d},\nonumber\\
L_{ij}&=&\left(
\begin{array}{ccc}
  \epsilon^{4} & \epsilon^{3} & \epsilon \\
  \epsilon^{3} & \epsilon^{2} & 1 \\
  \epsilon^{3} & \epsilon^{2} & 1 \\
\end{array}
\right)\epsilon^{p}H_{d}, ~~~~~ \nu^D_{ij}=\left(
\begin{array}{ccc}
  \epsilon^{2} & \epsilon & \epsilon \\
  \epsilon & 1 & 1 \\
  \epsilon & 1 & 1 \\
\end{array}
\right)\epsilon^{a_1}H_{u},
\end{eqnarray}
where $U_{ij}$, $D_{ij}$, $ L_{ij}$ and $\nu^D_{ij}$ correspond to
the up-quark, down quark, charged lepton and Dirac neutrino Yukawa
matrices resulting from the appropriate powers of the  $S$ field
in Eq. (\ref{fnn}). The integer $p$ can be either  0, 1 or 2,
corresponding to large, medium and small $\tan \beta\equiv \left<H_u\right>/\left<H_d\right>/$
respectively. Notice that the down-type quark mass matrix and the charged lepton
mass matrix are transpose of each other as required by an embedding into an $SU(5)$ GUT.

Once the charged lepton sector and Dirac neutrino sector are
constructed, we can uniquely define the form of the heavy Majorana
neutrino mass matrix.  In the present example it is
 \beq
 \label{neu4}
\nu_{ij}^M=M_R\left(
\begin{array}{ccc}
  \epsilon^{2} & \epsilon & \epsilon \\
  \epsilon & 1 & 1 \\
  \epsilon & 1 & 1 \\
\end{array}
\right)\epsilon^{a_2}~. \eeq
Any $SU(5)$ compatible theory automatically satisfies the
GS anomaly cancellation mechanism. This structure can naturally arise from the anomalous $U(1)_A$ type model.
However, as indicated earlier,the anomalous $U(1)_A$ is broken, we then identify discrete subgroups of the anomalous $U(1)_A$
symmetry as the discrete flavor gauge symmetry.

Three examples of $Z_{14}$ symmetric models are presented in Table
\ref{z14}.  We have chosen the charge of $S$ to be 2 and fixed the charge
of $\theta$ to be 7 in these examples.  Discrete anomaly
cancellation is enforced via GS mechanism at Kac--Moody level 1.
We have also imposed the conditions that the $Z_{14}$ symmetry
forbid all $R$-parity violating couplings.

\begin{table}\label{z14}
{\small
 \begin{center}
  {\renewcommand{\arraystretch}{1.1}
  \begin{tabular}{ c c c c  c  c  c  c  c  c  cc  c }
     \hline
   \rule[5mm]{0mm}{0pt}
  &$Q_i$&$u^{c}_i$&
   $d^{c}_i$&$L_i$&
   $e^{c}_i$&$\nu^{c}_i$ &
   $H_{u}$&$H_{d}$&
   $\theta$&$S$&$A_2$&$A_3$\\
   \hline
   \rule[5mm]{0mm}{0pt}
    A&0,2,6&1,3,7&3,5,5&4,6,6&13,1,5&5,7,7&1&13&7&2 &6& 13\\
    \rule[5mm]{0mm}{0pt}
    B&4,6,10&13,1,5&11,13,13& 6,8,8&9,11,1&5,7,7&13&1&7&2&13&13\\
    \rule[5mm]{0mm}{0pt}
    C&6,8,12&5,7,11&1,3,3&0,2,2&7,9,13&5,7,7&9&5&7&2&13&6\\
   \hline
  \end{tabular}
  }
  \caption{ Examples of the flavor--dependent $Z_{14}$ symmetry which
  forbids all $R-$parity breaking terms.
   $i=1,2,3$ is the flavor index and charges are in order of 1-3.
   }

 \end{center}}
\end{table}
We are considering $p=2$
   and $q=0$ in Eq. (\ref{matr1}) which  corresponds to medium
   values of $\tan\beta\sim 10$. We have taken $a_2=0$ in Eq. (\ref{neu4})
   for simplicity.

The above discrete gauge symmetries are consistent with realistic structure of fermion masses hierarchy in \ref{ckm} and \ref{ckm2}. And at the same time,
it gives the large mixing of neutrinos $\nu_\mu$ and $\nu_\tau$. Moreover, as discrete gauge symmetries, the famous $D$-term splitting problem can be then avoided.

\chapter{\textsc{Stabilization of Axion Solutions}}

\section{Strong CP Problem and QCD Axion}
CP violation (CPV) can exist in the QCD Lagrangian arising from
the instanton induced Chern-Simons type gluon-gluon coupling \beq
{\mathcal L}\supset\theta g^2_s
\epsilon^{\mu\nu\sigma\rho}G^{\alpha}_{\mu \nu}
G^{\alpha}_{\sigma\rho}/{64\pi^2}=\theta g^2_s G^{\alpha}_{\mu
\nu} \tilde{G}^{\alpha\mu\nu}/{32\pi^2}.\eeq In addition, there is
another CPV source from the quark mass matrices. This results in
an observable parameter $\bar{\theta}$ defined as \beq
\bar{\theta}=\theta+\mathrm{arg}(\mathrm{det}M_U~
\mathrm{det}M_D).\eeq Such a $\bar{\theta}$ would lead to a
neutron electric dipole moment (EDM) of order $d_n\simeq
5\times10^{-16}~\bar{\theta}~\mathrm{e cm}$, while the current
experiment limit is $d_n<10^{-25}~\mathrm{e cm}$. This puts a
strong constraint, $\bar{\theta}< 10^{-10}$. The PQ
symmetry \cite{PQ} is an elegant solution to this so-called strong
CP problem. It introduces a global $U(1)$ symmetry, broken by the
QCD anomaly, which generates a pseudo-Goldstone particle $a$, the
axion. Non-perturbative effects then induce a term in the
lagrangian of the form \beq {\mathcal L}\supset (a/f_a) g^2_s G^{\alpha}_{\mu
\nu} \tilde{G}^{\alpha\mu\nu}/{32\pi^2}.\eeq $\overline{\theta}$
is then promoted to this dynamical field axion as $a(x)/f_a$.
Minimizing the axion potential \beq  V(a) \propto
\Lambda^4_\rm{ QCD}(1-\cos(a(x)/f_a)),\eeq consequently
$\bar{\theta}=\langle a\rangle/f_a=0$.\footnote{Due to the
periodicity of the potential, $\langle a\rangle=2n\pi f_a$. Some
detailed discussion can be found in various review papers listed
in Ref. \cite{wk}.} The strong CP problem is then
solved.
 $f_a$
is the (model dependent) axion decay constant \cite{WW} and it is
constrained to be $f_a=(10^{10}-10^{12})~\rm{GeV}$ by the
combined limits from laboratory experiments, astrophysics and
cosmology. Hence, only the ``invisible axion" models, which have
appropriate values of $f_a$, are favored \cite{DFSZ,KSVZ}. The
couplings of the axion with the SM fields are
highly suppressed in these models. Although the axion arise as a
pseudo-Goldstone particle when the PQ symmetry is explicitly
broken by its QCD anomaly, the axion can acquire a tiny mass
through higher order non-perturbative effects. The mass of the
axion can be estimated to be \beq m_a\sim \Lambda^2_\rm{QCD}/f_a\sim 10^{-4}~\rm{eV}.\eeq

\section{Discrete Gauge Symmetry Stabilizing the Axion}
Quantum gravitational effects can potentially violate the global
PQ symmetry as they can break all global symmetries while
respecting gauge symmetries. In the axion models, a possible
quantum gravity generated non-renormalizable term \beq {\mathcal
L}\supset S^n/M_\rm{ Pl}^{n-4}\eeq is in principle allowed.
This term would lead to \beq \bar{\theta} \simeq
f_a^{n}/({{M_\rm{ Pl}}^{n-4}\Lambda_{QCD}^{4}}).\eeq Since
both $\bar{\theta}$ and $f_a$ are highly constrained, $n \geq 10$
is necessary. To avoid such kind of violations, one solution is to
introduce a discrete gauge symmetry \cite{kw}. The PQ symmetry  arises only
as an accidental global symmetry from it.

Conventionally, absence of anomalies complicates the particle
spectrum of axion models. However, the Type I and Type IIB string
theories provide a new candidate that cancels the anomalies
without enlarging the particle content. In the low energy
effective theory of such string theories, there exists one
anomalous $U(1)_A$ symmetry as mentioned in Section 4.3 \cite{gs,dsw}.
There, GS mechanism is effective in cancelling the anomalies. The
anomalous $U(1)_A$ symmetry is broken by a Higgs field
spontaneously near the $M_\rm{St}$. A discrete version of GS mechanism as in the Eq. \ref{dgsm} is applied here.

\section{Stabilization of the DFSZ Axion}

The non-SUSY DFSZ axion model \cite{DFSZ} introduces two Higgs doublets
$H_u$ and $H_d$ and a SM singlet scalar $S$.  The Lagrangian
of the model relevant for the discussion of axion physics is
\beq \label{ds1}
{\mathcal L} = Qu^{c}H_u+Qd^{c}H_d+Le^{c}H_d+\ell\nu^{c}H_u+\nu^{c}\nu^{c}-
\lambda({H_u}H_{d}S^{2}+\rm{h. c.}).\eeq
Here we have used a standard notation that easily generalizes to
our SUSY extension as well.

The ${\mathcal L}$ has three $U(1)$ symmetries, as can be inferred by solving
the six conditions imposed on nine parameters.
These three $U(1)$ symmetries can be identified as the SM hypercharge $U(1)_Y$,
baryon number $U(1)_B$ and a PQ
symmetry $U(1)_{PQ}$.
If we denote the charges of $(Q$, $u^c$,
$d^c)$ as $(q,~u,~d)$, the symmetries can be realized as $B=q-u-d$,
$PQ=-d$, $Y/2=q/6-2u/3+d/3$.  The $U(1)$ charges of the various
particles under these symmetries are listed in the Table \ref{pqc}.

\begin{table}[ht]\label{pqc}
 \begin{center}
  {\renewcommand{\arraystretch}{1.1}
  \begin{tabular}{ c  c  c  c  c  c  c  c  c  c  }
     \hline

     \rule[5mm]{0mm}{0pt}
   & $Q$ & $u^{c}$ &
   $d^{c}$ & $\ell$ &
   $e^{c}$ & $\nu^{c}$ &
   $H_d$ & $H_u$ &
   $S$\\
   \hline
   \rule[5mm]{0mm}{0pt}
   Y/2 & 1/6 & $-2/3$ & 1/3 &
   $-1/2$ & 1 & 0 & $-1/2$ & 1/2 & 0\\
    \rule[5mm]{0mm}{0pt}
    $B$ & 1 & $-1$ & $-1$ & 0 & 0 & 0 & 0 & 0 & 0\\
    \rule[5mm]{0mm}{0pt}
   $PQ$ & 0 & 0 & $-1$ &
   0 & $-1$ & 0 & 1 & 0 & $-1/2$\\
   \hline
  \end{tabular}
  }
  \caption {$Y/2$,
  $B$ and $PQ$ symmetries corresponding to hypercharge, baryon number and PQ
  charge respectively. The charges are assumed to be generation independent.}
   \end{center}
\end{table}

After $H_d$, $H_u$ and $S$ fields develop VEVs, the global PQ symmetry is
broken and the light spectrum contains a  Goldstone boson, the
axion. Non-perturbative QCD effects induce an
axion mass \cite{kim} given by
\beq
\label{ax1}
 m^{DSFZ}_a \simeq 0.6 \times 10^{-4}~ \rm{ eV}~ \frac {10^{11}~
 \mathrm{GeV}}{f_a},
 \eeq
where $f_a \sim \left\langle S \right\rangle$ is the axion decay constant.

We now apply the GS mechanism for discrete anomaly cancellation
to stabilize the axion from quantum gravity corrections.  Even though the
model under discussion is non-SUSY, the GS mechanism for
anomaly cancellation should still be available, since SUSY breaking
in superstring theory need not occur at the weak scale in principle.
Since baryon number has no QCD anomaly, any of its subgroup will be
insufficient to solve the strong CP problem via the PQ mechanism.
On the other hand, the PQ symmetry does have a QCD anomaly, although with the
charges listed in Table 1 it has no $SU(2)_L$ anomaly.  Since hypercharge
$Y$ is anomaly free, we attempt to identify the anomalous $U(1)_A$ symmetry as
a linear combination of $PQ$ and $B$:
\beq
 \label{an1} U(1)_A=PQ + \gamma B.
 \eeq
According to the Eq. (\ref{an1}) and the charge assignment presented
in Table 1, we have for the anomaly coefficients for the $U(1)_A$,
\begin{eqnarray}
 \label{an2}
 A_3&\equiv & [SU(3)]^2\times U(1)_A=-\frac{3}{2}  \nonumber\\\
 A_2&\equiv & [SU(2)]^2\times U(1)_A= \frac{9}{2}~\gamma~.
\end{eqnarray}
If we identify $\gamma = -k_2/(3k_2)$, the anomalies in $U(1)_A$
will be cancelled by GS mechanism.  Thus we have
 \beq \label{an4}
U(1)_A=PQ-\frac{1}{3}\frac{k_2}{k_3}B. \eeq
The simplest possibility is $k_2=k_3 =1$, corresponding to the levels of
Kac-Moody algebra being one.
Normalizing the charge of the singlet field $S$ to be an integer,
Eq. (\ref{an4}) can be rewritten as
 \beq \label{an5}
U(1)_A=6 (PQ)-2(B). \eeq
The corresponding charge assignment is given in Table
\ref{charge5}. As discussed earlier, since hypercharge $Y$ is
anomaly free, one can add a constant multiple of $Y/2$ to the
$U(1)_A$ charges, and still realize GS anomaly cancellation mechanism.
The charges listed in Table 6.2 assumes the combination
$-\frac{5}{3}(6PQ-2B+\frac{4}{5}Y)$.  As can be seen from Table 2,
this choice of charges is compatible with $SU(5)$ grand unification.

Suppose that the $U(1)_A$ symmetry is broken near the string scale
by the VEV of a scalar field which has a $U(1)_A$ charge of $N$
in a normalization where all $U(1)_A$ charges have been made integers.
A $Z_N$  symmetry will then be left unbroken to low scales.  Two
examples of such $Z_N$ symmetries are displayed in Table 2 for $N= 11,12$.
Invariance under these $Z_N$ symmetries will not be spoiled by quantum
gravity, it is this property that we use to stabilize the axion.

Potentially dangerous terms that violate the $U(1)_{PQ}$ symmetry
are $S^n/M_{\rm P\rm l}^{n-3}$, $H_u H_d S^{* m}/M_{\rm P\rm l}^{m-2}$
etc, for positive integers $n,m$.  For the induced $\bar{\theta}$ to be
less than $10^{-10}$, the integers $n,m$ must obey $n \geq 10, m \geq 5$.
The choice of $N=11,12$ satisfy these constraints.  Note that  a $Z_{10}$
discrete symmetry would have allowed a term $S^2$, which would be
inconsistent with the limit on $\bar{\theta}$.  $Z_N$ symmetries with $N$ larger than 12
can also provide consistent solutions. Since by construction, the $U(1)_A$
symmetry in Table 1 is anomaly-free by GS mechanism,
any of its $Z_N$ subgroup is also anomaly-free by the discrete GS mechanism,
as can be checked directly.  In the $Z_{11}$ model, for example, we have
$A_3 = A_2 = 4$.  Consistent with the $Z_{11}$ invariance, terms that
violate the  $U(1)_{PQ}$ symmetry and give rise to an axion mass are
$S^{11}/M_{\rm P\rm l}^7,~H_uH_dS^{*9}/M_{\rm P\rm l}^7$ etc, all of which are quite harmless.
We conclude that the DFSZ axion
can be stabilized against potentially dangerous non-renormalizable
terms arising from quantum gravitational effects in a simple way.

\begin{table}[h]
 \begin{center}
  {\renewcommand{\arraystretch}{1.1}
  \begin{tabular}{ c  c  c  c  c  c  c  c  c  c  }
     \hline
     \rule[5mm]{0mm}{0pt}
   & $Q$ & $u^{c}$ &
   $d^{c}$ & $\ell$ &
   $e^{c}$ & $\nu^{c}$ &
   $H_u$ & $H_d$ &
   $S$ \\
   \hline
   \rule[5mm]{0mm}{0pt}
$U(1)_A$& 2 & 2 & 4 &
   4 & 2 & 0 & $-4$ & $-6$ & 5 \\
   \rule[5mm]{0mm}{0pt}
$Z_{11}$& 2 & 2 & 4 &
   4 & 2 & 0 & 7 & 5 & 5 \\
\rule[5mm]{0mm}{0pt}
$Z_{12}$& 2 & 2 & 4 &
   4 & 2 & 0 & 8 & 6 & 5 \\
   \hline
  \end{tabular}
  }
  \caption{ The anomalous $U(1)$ charge assignment for
  the DFSZ axion model.  Also shown are the charges under two discrete subgroups
  $Z_{11}$ and $Z_{12}$ which can stabilize the axion.}
  \label{charge5}
 \end{center}
\end{table}

The discussion can be easily extended to its SUSY version.
The superpotential of the DFSZ axion model
contains a term $\lambda {H_u}H_{d}S^{2}/M_\rm{ Pl}$. After
$H_u$, $H_d$ and $S$ develop VEVs, the global PQ symmetry is
broken and the axion arises as a pseudo-Goldstone particle. Since
the superpotential is holomorphic, one cannot write ${S^\dagger}^2
S^2$ type term. In addition to the $S$ field, another singlet
$\tilde{S}$ is needed so that the axion is invisible and at the
same time, PQ can be broken. The superpotential of the model now
is \beq W\supset \lambda_1 H_u H_d S^2/M_\rm{ Pl}+ \lambda_2
S^2 \tilde{S}^2/M_\rm{ Pl}.\eeq One explicit example of
$Z_{22}$ discrete gauge symmetry is given. The charge assignment
under $Z_{22}$ is listed as \beq \{Q=3, ~u^{c}=19, ~d^{c}=1,
~\ell=11, ~e^{c}=15, ~\nu^{c}=11, ~H_u=22, ~H_d=18, ~S=13,
~\tilde{S}=20 \}.\eeq The mixed anomalies are  $\{A_2=6,
   A_3=17\}$. It apparently satisfies the GSM condition. $S^{22}/M^{19}_\rm{ Pl}$ is the leading allowed term in the
superpotential due to potential quantum gravity correction, which
only induces $\bar{\theta}\lesssim 10^{-130}$.

In this model, the $R$-parity is not automatic, for instance, $L
H_u S \tilde{S}$ is allowed. To get an exact $R$-parity, one can
introduce an additional $Z_2$ where all the SM matter fields are
odd but $H_u$, $H_d$, $S$ and $\tilde{S}$ are even. This is the
unbroken subgroup of the gauge symmetry $U(1)_\rm{ B-L}$ even
with the presence of Majarona neutrino mass term.

\section{Stabilization of KSVZ Axion}

The Kim-Shifman-Vainshtein-Zakharov (KSVZ) Axion model\cite{kim}, can also be stabilized by discrete
gauge symmetries. The scalar sector of the non-SUSY KSVZ axion model
\cite{KSVZ} contains the SM doublet and a singlet field $S$.  All
the SM fermions are assumed to have zero PQ charge under the
global $U(1)_{PQ}$ symmetry.  The Yukawa sector involving the SM
fermions is thus unchanged.  An exotic quark-antiquark pair $\Psi
+ \bar{\Psi}$ is introduced, which transform vectorially under the
SM (so the magnitude of its mass term  can be much larger than the
electroweak scale), but has chiral transformations under
$U(1)_{PQ}$.  The QCD anomaly needed for the axion potential
arises from these exotic quarks.  The Lagrangian involving the
singlet field and these vector quarks is given by
\beq \Delta {\mathcal L}=S\Psi\bar{\Psi}+ \rm{h.c.} \eeq
When $S$ field develops a VEV, the PQ symmetry is spontaneously
broken leading to the axion in the light spectrum.

The global PQ $U(1)$ symmetry is susceptible to unknown quantum
gravity corrections.  We shall attempt to stabilize the KSVZ axion
by making use of discrete gauge symmetries with anomaly
cancellation by the GS mechanism.  The most dangerous
non-renormalizable term in the scalar potential that can
destabilize the axion is $S^n/M_\rm{ Pl}^{n-4}$, as in the
case of the DFSZ axion.  We seek a discrete gauge symmetry that
would forbid such terms.

In order for the GS mechanism for anomaly cancellation
to be viable, the anomaly coefficients $A_2$ and $A_3$
corresponding to the $[SU(2)_L]^2 \times U(1)_A$ and $[SU(3)_C]^2
\times U(1)_A$ should equal each other at the $U(1)$ level.  This
would imply that the $\Psi + \bar{\Psi}$ fields can not all be
singlets of $SU(2)_L$. The simplest example we have found is the
addition of a $5+ \bar{5}$ of $SU(5)$ to the SM spectrum. Such a
modification is clearly compatible with grand unification. The
${\bf 5}$ contains a $({\bf3,1})$ and a $({\bf 1,2})$ under
$SU(3)_C \times SU(2)_L$.  We allow the following Yukawa coupling
involving these fields:
\begin{equation}
{\mathcal L} \supset \lambda {5} \bar{5} S + \rm{h.c.}
\end{equation}
If we denote the PQ charges of ${ 5}$ and ${ \bar{5}}$ as $\phi$
and $\bar{\phi}$, invariance under a surviving
discrete $Z_N$ symmetry would imply
\begin{equation}
\phi + \bar{\phi} + s = pN
\end{equation}
where $p$ is an integer.  In this simple model, all the SM
particles are assumed to be trivial under the PQ symmetry. The
discrete anomaly coefficients are then
$A_3=A_2=\frac{3}{2}(\phi+\bar{\phi})=\frac{3}{2}(p N-s)$.  Since
$A_2 = A_3$, the gauge anomalies are cancelled by the GS
mechanism. As long as $N \geq 10$, all dangerous couplings that
would destabilize the axion through non-renormalizable terms will
be sufficiently small.  We see that the KSVZ axion can be made
consistent in a simple way.

We have also examined the possibility of stabilizing the axion by
introducing only a single pair of fermions under the SM gauge
group, rather than under the grand unified group. Let us consider
a class of models with a pair of fermions transforming under
$G_\rm{SM}\times U(1)_A$ as
\beq \Psi({\bf
3,n,}y,\psi)+\bar{\Psi}(\bar{3},\bar{n},-y,\bar{\psi})~, \eeq
along with a scalar field $S({1,1},0,s)$. The Lagrangian of this
model contains a term $\Psi \bar{\Psi}S$  and its
invariance under an unbroken $Z_N$ symmetry imposes the constraint
\beq \psi+\bar{\psi}+s=p N\eeq where $p$ and $N$ are integers.
Since the SM particles all have zero anomalous $U(1)$ charge, the
anomaly coefficients arise solely from the $(\Psi+\bar{\Psi})$
fields.  They are
\begin{eqnarray}
A_3&=&\frac{1}{2}(n\psi+n\bar{\psi})=\frac{n}{2}(pN-s)\nonumber\\
A_2&=&\frac{(n-1)n(n+1)}{12}(3\psi+3\bar{\psi})\nonumber\\
&=&\frac{(n-1)n(n+1)}{4}(pN-s).
\end{eqnarray}
The GS discrete anomaly cancellation condition implies \beq s=pN+\frac{2(-m+bm')}{n(b(n^2-1)-2)}\eeq where
$b \equiv k_3/k_2$.

By choosing specific values of the Kac-Moody levels, one can
solve for $s$, the singlet charge. For instance, in the simple
case when $k_3=k_2\Leftrightarrow b=1$, \beq s=
\frac{2(m'-m)}{n^3-3n}N . \eeq We  have normalized all $U(1)_A$
charges to be integers, including $s$, so the unbroken $Z_N$
symmetry will be transparent.

When $n=2$, $\Psi$ and $\bar{\Psi}$ are $SU(2)$ doublets. One can
calculate the charge of $S$ and determine the
allowed discrete symmetries. For $b=1$, the solution is  $s=0~mod~
N$.  This solution would imply that $S^n$ terms in the potential
are allowed for any $n$, in conflict with the axion solution.  A
similar conclusion can be arrived at for $b = 1/2$.  For other
values of $b$, the $Z_N$ symmetry typically turns out to be too
small to solve the strong CP problem.  For example, if
$b=(2,3,1/3,3/2)$, the allowed discrete symmetries are $(Z_4, Z_7,
Z_3, Z_5)$.  A special case occurs when $b= 2/3$, in which case
$s$ is undetermined, since $A_3/k_3 = A_2/k_2$. If one chooses $s
\geq 10$, the KSVZ axion can be stabilized in this case.

If the quarks $\Psi$ and $\bar{\Psi}$ are triplets of $SU(2)_L$,
stability of the KSVZ axion solution can be guaranteed in a simple
way. For $b\equiv k_3/k_2 = (1,2,3,1/2,1/3,2/3,3/2)$, which are
the allowed possibilities if we confine to Kac-Moody levels less
than 3, we have the unbroken discrete symmetries to
be $(Z_9, Z_{21}, Z_{33}, Z_6, Z_3, Z_{15}, Z_{30})$ respectively.
For all $Z_N$ with $N \geq 10$, the axion solution will be stable
against quantum gravitational corrections.


\chapter{Conclusions}
In this thesis, we study the discrete gauge symmetries in the SM and also as a model building tool to solve various problems
in the SM as well as the MSSM.

In the second chapter, we discuss a hidden discrete gauge symmetry in the non-SUSY flavor independent SM at
the renormalizable level. A discrete $Z_3$ symmetry is found in the SM and is embedded into a discrete $Z_6$
symmetry in the extension of the SM with seesaw mechanism for the small neutrino masses. Both $Z_3$ and $Z_6$ are
free from mixed $G_\rm{SM}$ anomalies at the discrete level. It is anomaly free as a result of the
existence of three generations ($N_g=3$). The symmetry can effectively act as the baryon number up
to the $\Delta B=3~\rm{mod}~3$ level which is also consistent with the prediction from non-perturbative corrections
in the Standard Model, such as electroweak instanton  and sphaleron processes.

Quantum mechanically, we estimate the triple nucleon decay rate which is predicted by the existence of this symmetry.
It turns out, that as a result of baryon parity, the current bounds on the proton lifetime show that the cutoff scale in
4D can be as low as ${\mathcal O}(10^2~\rm{GeV})$.

We also find a simple $U(1)$ realization from which this baryon parity can naturally emerge. It is a $U(1)$ of
$I^3_R+L_i+L_j-2L_k$, where $I^3_R$ is the lepton number.

Effects arising from simple GUTs will explicitly break the baryon parity. Hence, whether there exists a
baryon parity puts a strong hint to GUT physics.

In Chapter 3, gauged $R$-parity is studied. It is shown that a $Z_2$ subgroup of $I^3_R$ plays an important role as
$R$-parity. After shifting the charges by a hypercharge rotation, one can realize this from a $Z_6$ subgroup
of the $U(1)_{B-L}$ symmetry.

In the forth chapter, we study the different approaches to the $\mu$-term problem, one puzzle in supersymmetric model building,
via a symmetry classification. Discrete gauge symmetries from the anomalous $U(1)_A$ symmetry can be applied to
solve this problem. One explicit example in terms of a  $Z_4$ symmetry is given, where the $\mu$-term problem is addressed
by the Giudice-Masiero mechanism. The SUSY DFSZ QCD axion is also discussed as other realization of the $\mu$-term problem.
Here, new physics scale $M_\rm{PQ}$ is imposed and $\mu$ arises as $\mu\sim M^2_\rm{PQ}/M_\rm{Pl}$.

Discrete flavor gauge symmetries are studied in the following Chapter 5 which can explain
the observed hierarchical structure of fermion masses while avoiding the $D$-term splitting problem in the usual SUSY soft
breaking sector. Discrete $Z_{14}$ gauged flavor symmetries are found to be consistent of one Lopsided hierarchical
structure of fermion masses.

In the last chapter, we show how to use discrete gauge symmetries to stabilize the ``invisible" axion solutions
from violation due to quantum gravity. The axion is an elegant solution to the strong CP problem.
Both DFSZ and KSVZ ``invisible axion" models are discussed. The PQ symmetry only arises as an accidental symmetry.
Examples of discrete $Z_{11}$ and $Z_{12}$ gauge symmetries are given to stabilize the non-SUSY DFSZ axion. For the SUSY
DFSZ case, a discrete $Z_{22}$ gauge symmetry is applied to stabilize the solution.

%
%
%

\bibitem{ua2}{
G.~Arnison \textit{ et al.}  [UA1 Collaboration],
Phys.\ Lett.\ B \bf{ 126}, 398 (1983);
P.~Bagnaia \textit{ et al.}  [UA2 Collaboration],
Phys.\ Lett.\ B \bf{ 129}, 130 (1983).}

\bibitem{lep}{The Electroweak Working Gorup, http://lepewwg.web.cern.ch/LEPEWWG/}

\bibitem{sm}{
S.~Weinberg,
Phys. Rev. Lett. \bf{ 19} (1967) 1264;
A. Salam, p.367 of \textit{ Elementary Particle Theory}, ed. N. Svartholm
(Almquist and Wiksells, Stockholm, 1969);
S.L. Glashow, J. Iliopoulos, and L. Maiani, Phys. Rev. D \bf{ 2} (1970) 1285.}


\bibitem{higgs}{
F. Englert and R. Brout, Phys. Rev. Lett. \bf{ 13} (1964) 321;
P.W. Higgs,
Phys. Lett. \bf{ 12} (1964) 132 and Phys. Rev. Lett. \bf{ 13} (1964) 508;
T.W. Kibble, Phys. Rev. \bf{ 155} (1967) 1554.}


\bibitem{solar}{
SNO Collaboration, Q.R. Ahmad \textit{ et al.},
Phys. Rev. Lett. \bf{ 87} (2001) 071301;
SNO Collaboration, Q. R. Ahmad \textit{ et al.}, { nucl-ex/0309004};
Super-Kamiokande Collaboration, S. Fukuda \textit{ et al.}, Phys. Lett. B
\bf{ 539} (2002) 179, { hep-ex/0205075}.}

\bibitem{atmospheric}{
Super-Kamiokande Collaboration, Y. Fukuda \textit{ et al.}, Phys. Rev. Lett.
\bf{ 81} (1998) 1562; Phys. Lett. B \bf{ 467} (1999) 185.}

\bibitem{KamLAND}{
KamLAND Collaboration, K. Eguchi \textit{ et al.},
Phys. Rev. Lett. \bf{ 90} (2003) 021802,
{ hep-ex/0212021}.}

\bibitem{K2K}{
K2K Collaboration, M.H. Ahn \textit{ et al.}, Phys. Rev. Lett. \bf{ 90}
(2003) 041801, { hep-ex/0212007}.}


\bibitem{wittenanomaly}{
E. Witten, Nucl. Phys. \bf{B 202} (1982) 253.}

\bibitem{wk4}{
K.~Wang,
arXiv:hep-ph/0402052.
}

\bibitem{wk3}{
K.~S.~Babu, I.~Gogoladze and K.~Wang,
Phys.\ Lett.\ B \bf{ 570}, 32 (2003)
[arXiv:hep-ph/0306003].
}

\bibitem{wk2}{
K.~S.~Babu, I.~Gogoladze and K.~Wang,
Phys.\ Lett.\ B \bf{ 560}, 214 (2003)
[arXiv:hep-ph/0212339].
}

\bibitem{wk1}{
K.~S.~Babu, I.~Gogoladze and K.~Wang,
Nucl.\ Phys.\ B \bf{ 660}, 322 (2003)
[arXiv:hep-ph/0212245].
}

\bibitem{operators}{ S. Weinberg, Phys. Rev. Lett. \bf{ 43} (1979)
1566; F. Wilczek and A. Zee, Phys. Rev. Lett. \bf{ 43} (1979)
1571.}

\bibitem{pdg} {K.~Hagiwara {\textit et al.}  [Particle Data Group Collaboration],
Phys. Rev. \bf{ D 66} (2002) 010001.}

\bibitem{zee} {H.A. Weldon and
A. Zee,  Nucl. Phys.\bf{ B 173} (1980) 269.}

\bibitem{leung}  {K.S. Babu and C.N. Leung, Nucl. Phys. \bf{ B
619} (2001) 667.}

\bibitem{gut}{
H. Georgi and S. Glashow, Phys. Rev. Lett. \bf{ 32} (1974) 438;
H. Georgi, {\em Unified Gauge Theories}
in {\em Proceedings, Coral Gables 1975, Theories and Experiments
In High Energy Physics}, New York (1975).}

\bibitem{luo}
{
S. Dimopoulos, S. Raby and F. Wilczek, Phys. Rev. \bf{ D24} (1981) 1681;
S. Dimopoulos and H. Georgi, Nucl. Phys. \bf{ B193} (1981) 150.
N.~Sakai, Z. Phys. \bf{ C11} (1981) 153;
L.E.~Ibanez and G.G.~Ross, Phys. Lett. \bf{ B105} (1981) 439;
L.E.~Ibanez and G.G.~Ross, Phys. Lett. \bf{ B110} (1982) 215;
M.B.~Einhorn and D.R.T.~Jones, Nucl. Phys. \bf{ B196} (1982) 475;
W.J.~Marciano and G.~Senjanovic, Phys. Rev. \bf{ D25} (1982) 3092;
P.~Langacker, M.-X.~Luo, Phys. Rev. \bf{ D44} (1991) 817;
J.~Ellis, S.~Kelley and D.V.~Nanopoulos, Phys. Lett. \bf{ B260} (1991) 131;
U.~Amaldi, W.~de Boer and H.~Furstenau, Phys. Lett. \bf{ B260} (1991) 447.
}

\bibitem{gswbook}{
M.~B.~Green, J.~H.~Schwarz and E.~Witten,
``Superstring Theory. Vol. 2: Loop Amplitudes, Anomalies And Phenomenology.''
}

\bibitem{wegner}{
F.~Wegner, J. Math. Phys. \bf{ 12}, 2259 (1971).}

\bibitem{krauss}{
J.~Preskill and L.~M.~Krauss,
Nucl.\ Phys.\ B \bf{ 341}, 50 (1990).
}

\bibitem{thooft}{
G.~'t Hooft,
Phys.\ Rev.\ Lett.\  \bf{ 37}, 8 (1976).}

\bibitem{ross}{
H.~K.~Dreiner and G.~G.~Ross,
Nucl.\ Phys.\ B \bf{ 410}, 188 (1993)
[arXiv:hep-ph/9207221].}

\bibitem{add}{N. Arkani-Hamed, S. Dimopoulos and G. Dvali, Phys. Lett. \bf{ B
429} (1998) 263; Phys. Rev. \bf{ D 59} (1999) 086004;
I.Antoniadis, N. Arkani-Hamed, S. Dimopoulos and G. Dvali, Phys.
Lett. \bf{ B 436} (1998) 257.}

\bibitem{littlehiggs} { N. Arkani-Hamed, A. G. Cohen, H. Georgi, Phys. Lett. \bf{ B 513} (2001)
232.}

\bibitem{kw}{ L. Krauss and F. Wilczek, Phys. Rev. Lett.\bf{ 182}
(1989) 1221}

\bibitem{GQW}{
H. Georgi, H. Quinn and S. Weinberg, Phys. Rev. Lett. \bf{ 33} (1974) 451.}

\bibitem{ptww}{
J.~Preskill, S.~P.~Trivedi, F.~Wilczek and M.~B.~Wise,
Nucl.\ Phys.\ B \bf{ 363}, 207 (1991).
}

\bibitem{banksdine}{
T.~Banks and M.~Dine,
Phys.\ Rev.\ D \bf{ 45}, 1424 (1992)
[arXiv:hep-th/9109045].
}

\bibitem{ibanez3}{
L.~E.~Ibanez and G.~G.~Ross,
Phys.\ Lett.\ B \bf{ 260}, 291 (1991).
}

\bibitem{ibanez1}{
L.~E.~Ibanez and G.~G.~Ross,
Nucl.\ Phys.\ B \bf{ 368}, 3 (1992).
}

\bibitem{yanagida}{
K.~Kurosawa, N.~Maru and T.~Yanagida,
Phys.\ Lett.\ B \bf{ 512}, 203 (2001)
[arXiv:hep-ph/0105136].
}

\bibitem{shafi}{G.~Lazarides and Q.~Shafi,
Phys.\ Rev.\ D \bf{ 58}, 071702 (1998) [arXiv:hep-ph/9803397].}

\bibitem{anomaly}{W.~A.~Bardeen,
Phys.\ Rev.\  \bf{ 184}, 1848 (1969);
S.~L.~Adler and W.~A.~Bardeen,
Phys.\ Rev.\  \bf{ 182}, 1517 (1969);
J.~S.~Bell and R.~Jackiw,
Nuovo Cim.\ A \bf{ 60}, 47 (1969).
}

\bibitem{nmssm}{
J.~R.~Ellis, J.~F.~Gunion, H.~E.~Haber, L.~Roszkowski and F.~Zwirner,
Phys.\ Rev.\ D \bf{ 39}, 844 (1989).}

\bibitem{hawking}{ S.W. Hawking, Phys. Lett. \bf{ B 195} (1987)337;
 G.V. Lavrelashvili, V.A. Rubakov and P.G. Tinyakov,  JETP
Lett. \bf{ 46} (1987) 167; S. Giddings and A. Strominger, Nucl.
Phys. \bf{ B 306} (1988) 349;  Nucl. Phys. \bf{ B 321} (1989) 481;
L.F. Abbot and M. Wise,  Nucl. Phys. \bf{ B 325} (1989) 687; S.
Coleman and K. Lee,  Nucl. Phys. \bf{ B 329} (1989) 389; R.
Kallosh, A. Linde,
 D. Linde and L. Susskind,  Phys. Rev. \bf{ D 52} (1995) 912.}

\bibitem{rsymmetry}{
L.~J.~Hall and L.~Randall,
Phys.\ Rev.\ Lett.\  \bf{ 65}, 2939 (1990);
L.~J.~Hall, Y.~Nomura and A.~Pierce,
Phys.\ Lett.\ B \bf{ 538}, 359 (2002)
[arXiv:hep-ph/0204062].
}

\bibitem{gasser}{See for example  J. Gasser and H. Leutwyler, Phys. Rept. {\bf 87}, 77
(1982).}

\bibitem{lattice}{\bibitem{Bakker:2004tw}
B.~L.~G.~Bakker, A.~I.~Veselov and M.~A.~Zubkov,
Phys.\ Lett.\ B {\bf 583}, 379 (2004).}

\bibitem{dias} {
A.~G.~Dias, V.~Pleitez and M.~D.~Tonasse,
Phys.\ Rev.\ D {\bf 69}, 015007 (2004)
[arXiv:hep-ph/0210172]; A.~G.~Dias, C.~A.~de S. Pires and P.~S.~R.~da Silva,
Phys.\ Rev.\ D {\bf 68}, 115009 (2003)
[arXiv:hep-ph/0309058]; \bibitem{Dias:2002gg}
A.~G.~Dias, V.~Pleitez and M.~D.~Tonasse,
Phys.\ Rev.\ D {\bf 67}, 095008 (2003)
[arXiv:hep-ph/0211107];
K. Dimopoulos, et al., hep-ph/0303154.}

\bibitem{kim}  {For reviews see: J.E. Kim, Phys. Rep. \bf{ 150} (1987) 1;
 H.Y. Cheng, Phys. Rep. \bf{ 158} (1988) 1; M.S. Turner, Phys.
Rep. \bf{ 197} (1991) 67; G.G. Raffelt, Phys. Rep. \bf{
333} (2000) 593;
G. Gabadadze and M. Shifman, Int. J.
Mod. Phys. \bf{ A17} (2002) 3689.}

\bibitem{DFSZ} { A.R. Zhitnitskii, Sov. J. Nucl. Phys. \bf{ 31} (1980) 260;
M. Dine, W. Fischler, M. Srednicki, Phys. Lett. \bf{ B104} (1981)
199.}

\bibitem{KSVZ}  {J.E. Kim, Phys. Rev. Lett. \bf{ 43} (1979) 103;
M. Shifman, A. Vainshtein, V. Zakharov, Nucl. Phys. \bf{ B166}
(1980) 493.}

\bibitem{seesaw}{M. Gell-Mann, P. Ramond and R. Slansky, in \textit{ Supergravity},
Proceedings of the workshop, Stony Brook, New York, 1979, edited
by P. van Nieuwnehuizen and D. Freedman (North-Holland, Amsterdam,
1979), p.315; T. Yanagida, in \textit{ Proceedings of the Workshop on
the Unified Theories and Baryon Number in Universe}, Tsukuba,
Japan, 1979, edited by O. Sawada and A. Sugamoto (KEK Report No.
79-18), Tsukuba, 1979), 95; R.N. Mohapatra and G. Senjanovi\'c,
Phys. Rev. Lett. \bf{ 44} (1980) 912.}

\bibitem{babumohap}{
K. S. Babu and R. N. Mohapatra, Phys. Rev. Lett. \bf{ 74} (1995)
2418.}

\bibitem{axionmu}{ J.E. Kim and H.P. Nilles, Phys. Lett. \bf{ 128B} (1984) 150;
E.J. Chun, J.E. Kim and H.P. Nilles, Nucl. Phys. \bf{ B370} (1992) 105.}

\bibitem{axino}{ E.J. Chun, J.E. Kim and H.P. Nilles, Phys. Lett. \bf{ B287}
(1992) 123.}

\bibitem{add}{
N. Arkani-Hamed, S. Dimopoulos and G. Dvali, Phys. Lett. \bf{ B
429} (1998) 263; Phys. Rev. \bf{ D 59} (1999) 086004; I.
Antoniadis, N. Arkani-Hamed, S. Dimopoulos and G. Dvali, Phys.
Lett. \bf{ B 436} (1998) 257.}
\bibitem{littlehiggs}{  N. Arkani-Hamed, A. G. Cohen, H. Georgi, Phys. Lett. \bf{ B 513} (2001)
232.}

\bibitem{hawking}{ S.W. Hawking, Phys. Lett. \bf{ B 195} (1987)337;
 G.V. Lavrelashvili, V.A. Rubakov and P.G. Tinyakov,  JETP
Lett. \bf{ 46} (1987) 167; S. Giddings and A. Strominger, Nucl.
Phys. \bf{ B 306} (1988) 349;  Nucl. Phys. \bf{ B 321} (1989) 481;
L.F. Abbot and M. Wise,  Nucl. Phys. \bf{ B 325} (1989) 687; S.
Coleman and K. Lee,  Nucl. Phys. \bf{ B 329} (1989) 389; R.
Kallosh, A. Linde,
 D. Linde and L. Susskind,  Phys. Rev. \bf{ D 52} (1995) 912.
\bibitem{ibanez} L.E. Ib\'{a}\~{n}ez and G.G. Ross, Nucl.
Phys. \bf{ B 368} (1992) 3.}

\bibitem{carone}{ C.D. Carone and H. Murayama, Phys. Rev. \bf{ D
52}(1995) 484.}
\bibitem{mohapatra}{T. Appelquist, et al. Phys. Rev. Lett. \bf{ 87} (2001)
181802;C. Lee, Q. Shafi and  Z. Tavartkiladze Phys. Rev. \bf{ D
66} (2002) 055010; R.N. Mohapatra and A. P\'{e}rez-Lorenzana,
Phys. Rev. \bf{ D 67} (2003) 075015.}
\bibitem{babu}{ See for example: K.S. Babu and R.N. Mohapatra, Phys. Lett. \bf{ B 532} (2002)
77.}
\bibitem{ma}{ E. Ma, Phys. Lett. \bf{ B 433} (1998) 74.
\bibitem{jlqcd} S. Aoki, et al. (JLQCD Collaboration) Phys. Rev.
\bf{ D 62} (2000) 014506.}

\bibitem{double}{
G. Feinberg, M. Goldhaber and G. Steigman, Phys. Rev. \bf{ D 18}
(1978) 1602; R. Mohapatra and G. Senjanovi\'{c}, Phys. Rev. Lett.
\bf{ 49} (1982) 7; L. Arnellos and W. J. Marciano, Phys. Rev.
Lett. \bf{ 48} (1982) 21; J. Basecq and L. Wolfenstein, Nucl.
Phys. \bf{ B 224} (1983) 21; S.P. Misra and U. Sarkar, Phys. Rev.
\bf{ D 28} (1983) 249; W.M. Alberico, et al, Phys. Rev. \bf{ C 32}
(1985) 1722;  K. Benakli and S. Davidson, Phys. Rev. \bf{ D 60}
(1999) 025004; C.E. Carlson and C.D. Carone, Phys. Lett. \bf{ B
512} (2001) 121.}

\bibitem{gs}{
M.B. Green and J.H. Schwarz, Phys. Lett. \bf{ B149} (1984) 117;
Nucl. Phys. \bf{ B255} (1985), 93; M. Green, J. Schwarz and P.
West, $ibid$. \bf{ B254} (1985) 327.}
\bibitem{majoron} {Y. Chikashige, R.N. Mohapatra and R.D. Peccei, Phys. Lett. \bf{ B 98} (1981) 265.}
\bibitem{babumj} {I.Z. Rothstein, K.S. Babu and D. Seckel, Nucl. Phys. \bf{ B
403} (1993) 725.}
\bibitem{nucleonsythen} {See for example: \textit{ The Early
Universe}, E.W. Kolb and M.S. Turner, Addison-Wesley Publishing
Company,  (1990)}

\bibitem{maru}{ K. Kurosawa, N. Maru, T. Yanagida, Phys. Lett. \bf{ B512} (2001)
203.}

\bibitem{ma} { E. Ma, Mod. Phys. Lett. \bf{ A17} (2002)535;
Phys. Rev. Lett. \bf{ 89} (2002) 041801.}

\bibitem{gm}{ G. Guidice and A. Masiero, Phys. Lett. \bf{ B206}, 1480 (1988).}

\bibitem{ramond} {For a review see  P. Ramond  hep-ph/9808488.}

\bibitem{dsw} {M. Dine, N. Seiberg and E. Witten, Nul. Phys. \bf{ B289}
(1987) 589; J. Atick, L. Dixon and A. Sen, Nucl. Phys. \bf{ B292} (1987) 109.}

\bibitem{ginsparg} {P. Ginsparg, Phys. Lett. \bf{ B197} (1987) 139.}

\bibitem{mass} {P. Binetruy, P. Ramond, Phys. Lett. \bf{ B350} (1995)
49; P. Binetruy, S. Lavignac, P. Ramond, Nucl. Phys. \bf{ B477}
(1996) 353; Y. Nir, Phys. Lett. \bf{ B354} (1995) 107; Z.
Berezhiani, Z. Tavartkiladze, Phys. Lett. \bf{ B396} (1997) 150;
Phys. Lett. \bf{ B409} (1997) 220; Q. Shafi, Z. Tavartkiladze,
Phys. Lett. \bf{ B482} (2000) 145; Phys. Lett. \bf{ B451} (1999)
129; M. Gomez et. al., Phys. Rev. \bf{ D59} (1999) 116009; J.
Feng, Y. Nir, Phys. Rev. \bf{ D61} (2000) 113005; A. S. Joshipura,
R. Vaidya and S. K. Vempati, Phys. Rev. \bf{ D62} (2000) 093020;
N. Maekawa, Prog. Theor. Phys. \bf{ 106} (2001) 401; I. Gogoladze,
A. Perez-Lorenzana, Phys. Rev. \bf{ D65} (2002) 095011; T.
Ohlsson, G. Seidl, Nucl. Phys. \bf{ B643} (2002) 247.}

\bibitem{dvali1}{ G. Dvali and S. Pokorski, Phys. Rev. Lett.\bf{ 78} (1997)
807; J.L. Chkareuli, C.D. Froggatt, I.G. Gogoladze, A.B.
Kobakhidze, Nucl. Phys. \bf{ B594} (2001) 23; Q. Shafi and  Z.
Tavartkiladze, Nucl. Phys. \bf{ B573} (2000) 40; N. Maekawa and T.
Yamashita, Prog. Theor. Phys.\bf{ 107} (2002) 1201;  B. Bajc, I.
Gogoladze, R. Guevara, G. Senjanovic, Phys. Lett. \bf{  B525}
(2002)189.}

\bibitem{cp} { K.S. Babu and S.M. Barr, Phys. Lett. \bf{ B300} (1993) 367; J. L. Lopez and D.V.
Nanopoulos, Phys. Lett. \bf{ B245} (1990) 111; K.S. Babu, B. Dutta and
R.N.Mohapatra, Phys. Rev. \bf{ D65} (2002) 016005.}

\bibitem{nilles}{ E.J. Chun, J.E. Kim and H.P, Nilles, Nucl. Phys.
\bf{ B370} (1992) 105; K. Choi, E.J. Chun and H.D. Kim,
Phys. Rev. \bf{ D55} (1997) 7010.}

\bibitem{dvali2} {G. Dvali and A. Pomarol, Phys. Rev. Lett.\bf{ 77} (1997)
3728; P. Binetruy, and E. Dudas, Phys. Lett. \bf{ B389} (1996)
503; R.N. Mohapatra and  A. Riotto, Phys. Rev. \bf{ D55} (1997)
4262.}

\bibitem{banksdine2}{
T.~Banks, M.~Dine and M.~Graesser,
Phys.\ Rev.\ D {\bf 68}, 075011 (2003)
[arXiv:hep-ph/0210256].}

\bibitem{martin} {R. N. Mohapatra, Phys. Rev. \bf{ D34} (1986) 3457; A. Font, L. Ibanez and F. Quevedo,
Phys. Lett. \bf{ B228} (1989) 79; S. Martin, Phys. Rev. \bf{ D46}
 (1992) 2769.}

\bibitem{dienes} {K. Dienes and J. March-Russell, Nucl. Phys. \bf{ B479}
(1996) 113;  K. Dienes, A. Faraggi and J. March-Russell, Nucl. Phys.
\bf{ B467} (1996) 44; S. Choudhuri, S.W. Chung, G. Hockney and
J. Lykken, Nucl. Phys. \bf{ B456} (1995) 89; Z. Kakushadze, G. Shiu,
S.H. Tye and Y. Vitorov-Karevsky, Int. J. Mod. Phys. \bf{ A13} (1998)
2551; K. R. Dienes, Phys. Rept. \bf{ 287} (1997) 447.}

\bibitem{fn} {C. Froggatt and H. Nielsen, Nucl. Phys. \bf{ B147} (1979)
277.}

\bibitem{PQ}{ R.~D.~Peccei and H.~R.~Quinn,
Phys.\ Rev.\ Lett.\  \bf{ 38}, 1440 (1977);
R.~D.~Peccei and H.~R.~Quinn,
Phys.\ Rev.\ D \bf{ 16}, 1791 (1977).
}

\bibitem{WW}  {S.~Weinberg,
Phys.\ Rev.\ Lett.\  \bf{ 40}, 223 (1978);
F.~Wilczek,
Phys.\ Rev.\ Lett.\  \bf{ 40}, 279 (1978).
}

\bibitem{axionmu}{ J.~E.~Kim and H.~P.~Nilles,
Phys.\ Lett.\ B \bf{ 138}, 150 (1984);
E.~J.~Chun, J.~E.~Kim and H.~P.~Nilles,
Nucl.\ Phys.\ B \bf{ 370}, 105 (1992).
}

\bibitem{muterm}{
J.~F.~Gunion and H.~E.~Haber,
Nucl.\ Phys.\ B \bf{ 272}, 1 (1986)
[Erratum-ibid.\ B \bf{ 402}, 567 (1993)]}

\bibitem{susyreview}{
H.~E.~Haber and G.~L.~Kane,
Phys.\ Rept.\  \bf{ 117}, 75 (1985); H.~P.~Nilles,
Phys.\ Rept.\  \bf{ 110}, 1 (1984); S.~P.~Martin,
arXiv:hep-ph/9709356.
}

\bibitem{u1p}{
M.~Cvetic, D.~A.~Demir, J.~R.~Espinosa, L.~L.~Everett and P.~Langacker,
Phys.\ Rev.\ D \bf{ 56}, 2861 (1997)
[Erratum-ibid.\ D \bf{ 58}, 119905 (1998)]
[arXiv:hep-ph/9703317].}

\bibitem{hierarchy}
{S.~Weinberg,
Phys.\ Rev.\ D \bf{ 13}, 974 (1976);
S.~Weinberg,
Phys.\ Rev.\ D \bf{ 19}, 1277 (1979);
L.~Susskind,
Phys.\ Rev.\ D \bf{ 20}, 2619 (1979);
G. 't Hooft, in Recent developments in gauge theories,
Proceedings of the NATO Advanced Summer Institute, Cargese 1979, ed. G. 't Hooft
et al. (Plenum, New York 1980).}

\bibitem{susy}{
E.~Witten,
Nucl.\ Phys.\ B \bf{ 188}, 513 (1981);
Z.\ Phys.\ C \bf{ 11}, 153 (1981);
S.~Dimopoulos and H.~Georgi,
Nucl.\ Phys.\ B \bf{ 193}, 150 (1981);
R.~K.~Kaul and P.~Majumdar,
Nucl.\ Phys.\ B \bf{ 199}, 36 (1982).}


\listrefsAIP
%
%
%
%
%
%
%
%
%
%

\thispagestyle{empty}
\topmargin -.575in
\singlespace
\noindent
\rm{Name:~~Kai Wang}\hfill \rm{Date of Degree:  July, 2004}

\vspace{12pt}
\noindent
Institution:~~Oklahoma State University\hfill Location:~~Stillwater, Oklahoma

\vspace{12pt}
\noindent
\rm{Title of Study:}\ \ \parbox[t]{4.5in}{HIDDEN SYMMETRIES AND THEIR IMPLICATIONS FOR PARTICLE PHYSICS}

\vspace{12pt}
\noindent
\rm{Pages in Study:~~61}\hfill \rm{Candidate for the
Degree of Master of Science}

\vspace{12pt}
\noindent
Major Field:~~Physics

\begin{description}
\item{Scope and Method of Study}:~~In this thesis, we study the role that discrete gauge symmetries
can play in solving phenomenological problems of the Standard Model and new physics beyond the Standard Model.
Discrete gauge symmetries were first introduced in the context of 4D particle physics as remnants of spontaneously broken
gauge symmetries by Krauss and Wilczek. Following their pioneering work, it
has been widely discussed to use discrete gauge symmetries as a new
model building tool. The key method here is to study the triangle gauge anomalies of the theories
 and how to cancel the  anomalies to have a gauge realization
of the symmetries.
\item{Findings and Conclusions}:~~Following the introduction part of the Chapter 1, the second chapter contains a discussion of so called ``gauged baryon parity",
a discrete gauge symmetry in the flavor independent non-supersymmetric
Standard Model at the renormalizable level. It is anomaly free as a result of the existence of three generations.
The symmetry can effectively act as the Baryon number up to the $\Delta B=3~\rm{mod}~3$ level which is also consistent with the
prediction from non-perturbative corrections in the Standard Model, such as electroweak instanton  and sphaleron processes.
Quantum mechanically, we estimate the triple nucleon decay rate which is predicted by the existence of this symmetry. We study the anomalies
of the symmetry and find a simple $U(1)$ realization from which this baryon parity can naturally emerge. New physics like GUTs
 will explicitly break the symmetry. In the third chapter, gauged $R$-parity is studied.
In the following Chapter 4, we study the different approaches to the $\mu$-term problem, one puzzle in supersymmetric model building, via a symmetry classification.
Discrete gauge symmetries
from the anomalous $U(1)_A$ symmetry have been used to solve this problem. One explicit example in terms of a  $Z_4$ symmetry is given.
Discrete flavor gauge symmetries
are studied in the following Chapter 5, which can explain
the observed hierarchical structure of fermion masses. The last chapter shows how to use discrete gauge symmetries to stabilize the axion solutions which are
usually regarded as
one elegant solution to the strong CP problem. Both DFSZ and KSVZ ``invisible axion" models are discussed.

\end{description}

\vspace{\fill}

\noindent
ADVISOR'S APPROVAL:\rule[-.1in]{4in}{.02in}
\doublespace

\end{document}